\documentclass[final,1p,times]{elsarticle}
\usepackage{amssymb}
\usepackage{amsmath}
\usepackage{amsthm}
\usepackage{tikz}
\usepackage{soul}

\biboptions{sort&compress}

\newcommand{\bs}[1]{{\boldsymbol{#1}}} %\newcommand{\bs}[1]{{\mathbf{#1}}}
\newcommand{\gr}[1]{( #1 )}
\newcommand{\Gr}[1]{\big( #1 \big)}
\newcommand{\GR}[1]{\Big( #1 \Big)}
\newcommand{\AutoGr}[1]{\left( #1 \right)}
\newcommand{\vect}[1]{[ #1 ]}
\newcommand{\set}[1]{\{ #1 \}}

\newcommand{\BigSet}[1]{\Big\{ #1 \Big\}}

\newcommand{\abs}[1]{\vert #1 \vert}

\newcommand{\norm}[1]{\Vert #1 \Vert}
\newcommand{\bin}[1]{\texttt{#1}}
\newcommand{\inflLOC}[2]{R_{#1}\gr{#2}} % \text{loc},
\newcommand{\RELinflLOC}[2]{r_{#1}\gr{#2}} % \text{loc},
\newcommand{\inflLEV}[2]{R_{\level{#1}}\gr{#2}}
\newcommand{\RELinflLEV}[2]{r_{\level{#1}}\gr{#2}}
\newcommand{\inflTOT}[1]{R\gr{#1}}
\newcommand{\RELinflTOT}[1]{r\gr{#1}}
\newcommand{\var}[1]{\sigma^2\gr{#1}}
\newcommand{\K}{s} % Legacy of K-sparse now s-sparse

\newcommand{\ith}{$i$th}
\newcommand{\jth}{$j$th}
\newcommand{\kth}{$k$th}
\newcommand{\Int}{\mathbb Z}

\newcommand{\Real}{\mathbb R}
\newcommand{\Syl}{H} % {\mathcal{H}}
\newcommand{\eps}{\varepsilon}
\newcommand{\level}[1]{\mathcal{L}_{#1}}  % \newcommand{\level}[2]{\mathcal{L}_{{#1},{#2}}}
\newcommand{\argmin}{\mathop{\mathrm{argmin}}}

\newdefinition{example}{\textbf{Example}}
\theoremstyle{definition}
\newtheorem*{remark}{Remark}

\journal{Journal of Theoretical Biology, {https://doi.org/10.1016/j.jtbi.2021.110985}}

\begin{document}

\begin{frontmatter}



\title{On the Fourier transform of a quantitative trait:\\ Implications for compressive sensing}
%\title{Discovering genetic networks using compressive sensing}


\author[1]{Stephen Doro\corref{cor1}} % \fnref{fn1}}
\ead{sd15@columbia.edu}
\cortext[cor1]{Corresponding author}
\address[1]{Columbia University, New York, NY, USA}

\author[2]{Matthew A. Herman}
\address[2]{Fourier Genetics, Austin, TX, USA}

\begin{abstract}
This paper explores the genotype-phenotype relationship. It outlines conditions under which the dependence of a quantitative trait on the genome might be predictable, based on measurement of a limited subset of genotypes. It uses the theory of real-valued Boolean functions in a systematic way to translate trait data into the Fourier domain. Important trait features, such as the roughness of the trait landscape or the modularity of a trait have a simple Fourier interpretation. Ruggedness at a gene location corresponds to high sensitivity to mutation, while a modular organization of gene activity reduces such sensitivity.

Traits where rugged loci are rare will naturally compress gene data in the Fourier domain, leading to a sparse representation of trait data, concentrated in identifiable, low-level coef\mbox{}ficients. This Fourier representation of a trait organizes epistasis in a form which is isometric to the trait data.  As Fourier matrices are known to be maximally incoherent with the standard basis, this permits employing compressive sensing techniques to work from data sets that are relatively small\,---\,sometimes even of polynomial size\,---\,compared to the exponentially large sets of possible genomes.

This theory provides a theoretical underpinning for systematic use of Boolean function machinery to dissect the dependency of a trait on the genome and environment.
\end{abstract}


\end{frontmatter}


\section{Introduction} \label{Sec:Intro}
% ==================================================================================
% ==============================================================
There is a practical problem in describing the relationship between the genome and the phenotype. Consider some quantifiable trait of an organism; examples include height, weight, antibiotic resistance, the spherical equivalent of the eye, and countless others. It is natural to seek to find the set of genes responsible for that trait. Genome-wide association studies (GWAS) can detect places in the chromosome with a strong statistical association with some trait. Alleles which influence the trait in question are located at such places, although the precision of GWAS may be insuf\mbox{}ficient to pinpoint the exact location responsible for the trait variability. However, even if the allele that correlates with a trait could be pinpointed, one cannot quantify the full ef\mbox{}fect of each gene on the trait in question, because the ef\mbox{}fect of one gene may be modified by the presence of other genes at other loci.

A na\"{i}ve approach would attempt to break the trait modification caused by each locus into a spectrum of ef\mbox{}fects; attach a number measuring the ef\mbox{}fect of swapping alleles at each of these loci (the ``allele ef\mbox{}fect size''~\cite{DistrAlleleFreqs_Park2011}) and then predict the resultant trait value by simply adding together the individual ef\mbox{}fects over the spectrum of loci. This approach is useful whenever the combined ef\mbox{}fects are suf\mbox{}ficiently close to linear.
The central limit theorem predicts that the values of a trait over a population with independently distributed alleles will converge to a Gaussian distribution~\cite[p.~$66$]{NaturalInheritance_Galton1889}. This theory underlies biometric analysis of ``heritability,'' which assumes that trait values are the convolution of genetic and environmental influences.

In practice, typical distributions of biological traits are routinely sketched as bell-shaped curves, even if these curves are not necessarily Gaussian. It is quite a dif\mbox{}ficult matter to rule out fat tails in limited data sets. The trait distribution need not converge to a Gaussian shape, when interaction terms between two or more loci are introduced. Of course, whenever interaction terms are sparse, we can expect a generally bell-shaped trait distribution. Such traits should be prime candidates for a compressive sensing analysis. The formalism to be presented in this paper permits arbitrary interactions and, therefore, can assume arbitrary trait distributions.  An analysis of the ef\mbox{}fect of binary interactions is discussed in~\cite[p.~399f]{EvolSelectQuantTraits_WalshLynch2018}. Even binary interactions can create leptokurtotic distributions.

But it is also quite clear that the na\"{i}ve analysis above is incorrect. Even in the simplest case of two homologous loci on paired chromosomes, examples of dominant and recessive gene interactions show that the ef\mbox{}fects of switching alleles at separate loci are not strictly additive. A trait is not a linear combination of individual and isolated gene ef\mbox{}fects\,---\,it emerges from activity and interactions \emph{within and across the entire genome}. For any particular trait, most of these genes contribute little to the trait variation and so may be relegated to a fixed background, but a complex trait will still require consideration of the interactions of a formidable number of loci. One can picture an underlying ``landscape'' of the various genomes associated with the trait, where the trait value is represented by the landscape elevation. But the geometry is not two-dimensional: it is discrete and multidimensional. To conclusively understand the ef\mbox{}fect of the genes, one would need a table assigning the value of the trait to every possible combination of the relevant genes at many locations. This is the so-called ``scale problem''~\cite{QuantAnalEmpFitnessLandscapes_Szendro2013}, so such a table would be enormous.

The number of possible genomes grows exponentially with the number of gene loci. For example, if we find $n$ loci (each with $2$ alternate alleles) associated with a certain trait, then there are $2^n$ genomes. To pick a relatively small case, if $n = 100$ there will be approximately $10^{30}$ potentially dif\mbox{}ferent genotypes. It would be flatly impossible to measure the trait for each such genotype and hence, the landscape function from genome to trait cannot be measured in any practical sense. The number of observations of the trait must be a small fraction of the possible genotypes.

William Bateson coined the term ``\emph{epistasis}'' for phenotypic alterations due to interactions between genes at separate loci~\cite{MendelsPrinciplesHeredity_Bateson1902, Bateson-BiologistAheadOfTime_Bateson2002}. Shortly afterwards, R.~A.~Fisher defined \emph{epistacy} in more mathematical terms as ``a deviation from the addition of superimposed ef\mbox{}fects \ldots between dif\mbox{}ferent Mendelian factors''~\cite[p.~$404$]{CorrelationBetweenRelatives_Fisher1918}. Moore and Williams~\cite{EpistasisImplicationsPersonalGenetics_moore2009} discuss the resultant tension between Bateson's biological and Fisher's mathematical epistasis and call for a more modern definition in the light of emerging knowledge of genetic networks.

Our approach is a specialization of Fisher's definition, which conjures up a hierarchy of epistatic interactions. There are various ways to organize such a collection~\cite{ContextDependenceMutations_Poelwijk2016}. Mathematically, the epistasis coef\mbox{}ficients comprise a non-singular (if it is not to lose information) linear transformation of the trait values. For many traits, Fisher's significant deviations from additivity may be rare enough\,---\,needles in the haystack of possible interactions\,---\,so that the epistasis transform may be discovered from observations considerably less than exponential, perhaps even of polynomial order.

This paper explores the sorts of traits and transforms which might permit a way around the scale problem. It is certainly not enough for a transform of trait data into epistasis coef\mbox{}ficients to be invertible. Ideally, the transform should compress the trait data into a sparse set of coef\mbox{}ficients; the rest being zero or negligibly small. Further, one would like the most important interactions to involve a relatively small number of gene loci, with complex interactions increasingly rare. It would also be desirable that the transform be a geometric similarity, so that changes of the trait produce proportionate changes of the transform, and vice versa. Finally, there is a rather technical desideratum, described more fully in Section~\ref{Sec:Compressive_Sensing}: the transform should be maximally ``incoherent.'' Loosely speaking, this means that each observation of a trait value provides some information about all epistasis entries.

These criteria coincide with the conditions (specifically, sparsity and incoherence) necessary to use compressive sensing to detect a manageable set of interactions from observations that are a small fraction of the exponentially vast space of genotypes. Our investigations rely on the well-established mathematical toolkit of real-valued Boolean functions~\cite{AnalysisBooleanFncs_ODonnell2014, NoiseSensBoolFncsPerc_Garban2014}. The biologist cannot be expected to be very familiar with this area of abstract algebra, and mathematical proofs would be out of place in this introduction, but a brief, impressionistic picture may whet the appetite.

The familiar Fourier transform maps functions on the real line to functions on a transform space; a sort of mirror space. Our discrete Fourier transform is ef\mbox{}fected by the Hadamard matrix. Just as in classical Fourier analysis, there is a linear transform from one space to the other, and the inverse transformation is (up to a scaling constant) the same as the direct. Trait functions now take the collection of bit-strings encoding the dif\mbox{}ferent genomes as an underlying basis. A more algebraic treatment would note that these bit-strings comprise an abelian group, leading to a generalized Fourier transform, but such an abstract treatment is not needed to understand this paper. This group, or collection of strings, may be represented geometrically by a hypercube. So, a quantitative trait may be pictured as a landscape (or more prosaically, a vector) on this multidimensional geometry and the trait's transform is a vector defined on a dif\mbox{}ferent hypercube. The specific coef\mbox{}ficients of the transform correspond to a particular version of epistasis~\cite{ContextDependenceMutations_Poelwijk2016}. We call the space of all possible traits the ``\emph{trait space}'' and the transform space the ``\emph{gene network space}'' because of its important analogies with other gene networks, such as the coexpression network and gene regulatory network.

This scheme expands the spectrum of gene ef\mbox{}fects from the individual loci to associate an interaction ef\mbox{}fect to every possible set, or \emph{cluster of loci} (e.g., as singletons, pairs, triads, etc.). Given this many interactions, any trait can be broken into a spectrum of interactions but there are now $2^n$ possibilities. The ``\emph{level}'' of a Fourier component indicates \emph{the number of gene loci in a given cluster}. It will turn out that the level-$0$ component, corresponding to the null set, is the average value of the trait, and the level-$1$ components, which correspond to a single locus, are simply the allele ef\mbox{}fect sizes, i.e., the change in the trait average from swapping allele `$\bin{0}$' to allele `$\bin{1}$', or vice versa, at a locus in question. Similarly, in the case of two homologous loci, the corresponding level-$2$ Fourier component encodes their interaction, with the sign expressing whether this is a dominant or a recessive interaction. In this context, the components associated with relatively few genes take the place of ``low-frequency'' terms in standard Fourier analysis, while ``high-frequency'' terms represent interactions involving many genes.

The Fourier transform acting on a quantitative trait may be viewed as an operator that \emph{untangles} the complicated relationships between genes expressed in the values of a trait. Hence, it makes visible many important features hidden in the trait values. The average value of a trait, the variance of a trait~\cite[p.~$399$]{CorrelationBetweenRelatives_Fisher1918}, the overall roughness of the trait terrain and the local contribution of individual loci to this roughness, all have simple expressions in terms of the trait's Fourier coef\mbox{}ficients. But crucially, we will argue the following. For
\begin{enumerate}
\item[(i)] highly evolvable traits, or \\[-10pt]
\item[(ii)] modular traits,
\end{enumerate}
\emph{the transform concentrates the trait information into low levels of the gene network, resulting in a sparse or compressible representation.} We say that a data set is \emph{$s$-sparse} if at most~$s$ of its elements are nonzero, and that it is \emph{compressible} if it is well-approximated as {$s$-sparse}. In addition to sparsity, the Fourier transform also provides the requisite incoherence to permit compressive sensing, so as to recover the important epistasis interactions from highly incomplete trait data. Then, an inverse transform applied to the gene network will reconstruct the entire trait, accurately predicting trait values for unobserved genotypes.

% ==============================================================
% ==================================================================================
\section{Overview of prior work}
% ==================================================================================
% ==============================================================
The relationship of genotype to phenotype is central to biology. Francis Galton and his school of biometricians developed tools to describe continuous traits, but in the early stages this outlook clashed with the discrete features of the genome explored by the Mendelians~\cite{provine2001origins}. Walsh and Lynch state, ``Vestiges of this dif\mbox{}ference between the gene-based focus of the Mendelians and the continuous-trait focus of the biometricians still persist today''~\cite[p.~$5$]{EvolSelectQuantTraits_WalshLynch2018}.

This split is finessed by the theoretical notion of a landscape: a real value trait defined over a discrete domain, often pictured with interpolated values, filling in the discrete domain~\cite{MeasuringRuggedness_VanCleve2015}. Like so much else, these graphs date back to Fisher~\cite{GeneticalTheory_Fisher1930}.
This construction is most frequently applied to the trait ``fitness,'' with the underlying domain being a hypercube, whose vertices are strings of zeros and ones~\cite{BeyondHypercube_Zagorski2016}, e.g., as in the NK~model~\cite{RuggedLandscapes_Kauffman1987, NKmodelRugged_Kauffman1989}, the Mount Fuji model~\cite{AnalysisLocalFitnessMtFuji_Aita2000}, and the ``holey'' landscape constructs~\cite{DynamicalTheoryHoley_Gavrilets1999}. The trait ``fitness'' may be defined in terms of success in creating of\mbox{}fspring, but is not readily measured. If the hypercube picture is extended to other, more measurable traits, the scale problem is evident.

Galton's work~\cite{NaturalInheritance_Galton1889} on the separation of hereditary and environmental influences depends on these influences being additive, but when Fisher introduced the notion of epistasis for non-linear interactions of distinct sets of genes, he did not pursue any specific scheme or normalization of the many possible epistasis coef\mbox{}ficients. It was eventually noted in the biological literature~\cite{EmpFitnessLandscapesPredEvol_deVisser_Krug2014} that a Fourier transform is one way to organize a trait's epistasis, but the connection of this observation to compressive sensing was not pursued until later~\cite{ContextDependenceMutations_Poelwijk2016}. If there are no limits on possible epistases, the scale problem prevents practical measurement of a trait. It has been shown that such epistasis are not confined to binary interactions~\cite{InferringShapeGlobalEpistasis_Otwinowski2018}.
Work has also recently been done detecting gene interactions using Fourier analysis on non-abelian groups~\cite{DetectGenomicSpectralAnalysis_Uminsky2019}.

Several methods have arisen to give some further structure to gene interactions, by fitting gene loci into a graph, such as gene coexpression networks or gene regulatory networks~\cite{StructureComplexNetworks_Estrada2011, GeneticNetworksFunctional_Pisabarro2008, InferringCellularNetworks_Markowetz2007, LargeNetworkSmallMolec_Sharom2004}: an edge in such networks should correspond to some non-linear interaction of two genes. The rise of computers has led to significant advances in the reconstruction of signals from incomplete data; compressive sensing and the related area of ``matrix completion'' have contributed to this progress. A number of approaches have been made to apply compressive sensing to gene networks and related data sensing problems~\cite{GeneRegNetworksCS_Chang2014, SparseEpistaticRegularizationDNN_Aghazadeh2020, HumanGeneCoexpressionLandscape_Prieto2008}; also see the relevant theoretical work in~\cite{Thesis_Stobbe2013}. In the case of compressive sensing with highly sparse Fourier domain signals, there is considerable overlap with the field of ``sparse fast Fourier transforms'' (sFFT)~\cite{sFFT_Indyk2014}.

A graph is a local structure, in some sense, since each edge represents a relationship of two gene loci. A more global structure is to break the loci into clusters. A body of work~\cite{SurvivalSparsestParsimonious_Leclerc2008, ModularityCostComplexity_WelchWaxman2003, MolecularModularCellBio_Hartwell1999, DevelopBasisVariationalMod_Mitteroecker2009, ModularityNecEvolvability_Hansen2003, EvolvabilityRobustness_2017, PleiotropyPreservationPerfection_Waxman1998} explores the organization of genes into separate functional units. Of course, the advantages of a modular arrangement are straightforward, but the question arises how such a universally designed feature could arise by the random mutations of evolution. A proposed solution is an evolutionary drift towards ``parsimony''~\cite{SurvivalSparsestParsimonious_Leclerc2008} or ``reduced complexity''~\cite{ModularityCostComplexity_WelchWaxman2003} or ``perfection''~\cite{PleiotropyPreservationPerfection_Waxman1998} whereby certain genes resist evolutionary change, because any alteration in such genes have multiple ef\mbox{}fects. In short, the landscape is ``rough'' at such genes and resists alteration. But a process of gene duplication and subsequent specialization leads to a smoothening of the trait landscape.

This makes it interesting to have a local measure of trait roughness and to examine the statistical distribution of roughness at the various loci. This paper has done this for one \emph{complete} data set, and the results suggest a power-law distribution consistent with that of evolving networks~\cite{StatMechComplexNetwork_Barabasi2002}, but of course, considerably more exemplars are needed.

The analysis of Boolean functions, and of real-valued Boolean functions is a foundational part of theoretical computer science. The key techniques revolve around the ability to read of\mbox{}f the properties of a Boolean function from its Fourier transform.

The marvelous work of Poelwijk, \emph{et al.}~\cite{ContextDependenceMutations_Poelwijk2016} shows an awareness that the epistasis coef\mbox{}ficients should be viewed as a linear transformation of the trait values. They give a variety of formulations of this transform, one of which is the Hadamard matrix, while others are scaled by powers of $2$, according to their level in the hierarchy. But these other \emph{ad hoc} transforms do not preserve the metric geometry of the transform. In a follow-up paper, Poelwijk, \emph{et al.}~\cite{LearningPatternEpistasis_Poelwijk2019} show that compressive sensing works admirably with the Hadamard matrix and fails for other transforms.

% ==============================================================
% ==================================================================================
\section{Traits and their associated gene networks} \label{Sec:Traits&theirNetworks}
% ==================================================================================
% ==============================================================
In common usage, a ``trait'' is any quantifiable aspect of an organism's phenotype, e.g.,  height, weight, etc. But the scale problem usually prevents us from observing the phenotype associated with all possible genotypes, so we will refer to a ``partial trait'' as the incomplete vector of trait values which are known to us, and to the ``full trait'' as a complete vector of trait values for every genotype. Ultimately, our mathematical analysis aims at reconstructing the full trait, from the information contained in a suf\mbox{}ficiently large partial trait, by exploiting hidden structure in the data.

In this paper, we are concerned with the mathematical dependence of traits on some set of binary alternatives (although this can easily be extended to variables with more than two states). Usually, these will be alternate gene alleles, but there are obvious extensions to the presence or absence of other factors, such as the environment or non-genetic comorbidities.

There is a close analogy between the well-developed theory of Boolean functions and a way of understanding the relationship between genotype and phenotype.\\

\textbf{Definition:} A \emph{trait} $\bs{t}$ is a real-valued function on a Boolean lattice of genotypes.\\

The Boolean lattice can be identified with the set of strings of zeros and ones of some fixed length $n$. The particular choice of `$\bin{0}$' or `$\bin{1}$' at some locus in this string encodes which allele is present at some corresponding point on the genome, or more generally, the presence or absence of some influencing agent. We will refer loosely to such a string as an artificial ``genome,'' a mathematical abstraction from the genuine genome.

\begin{figure}[!tb]
\centering
\includegraphics[scale=0.5, trim={70mm 47mm 70mm 45mm},clip] % {left, bottom, right, top}
{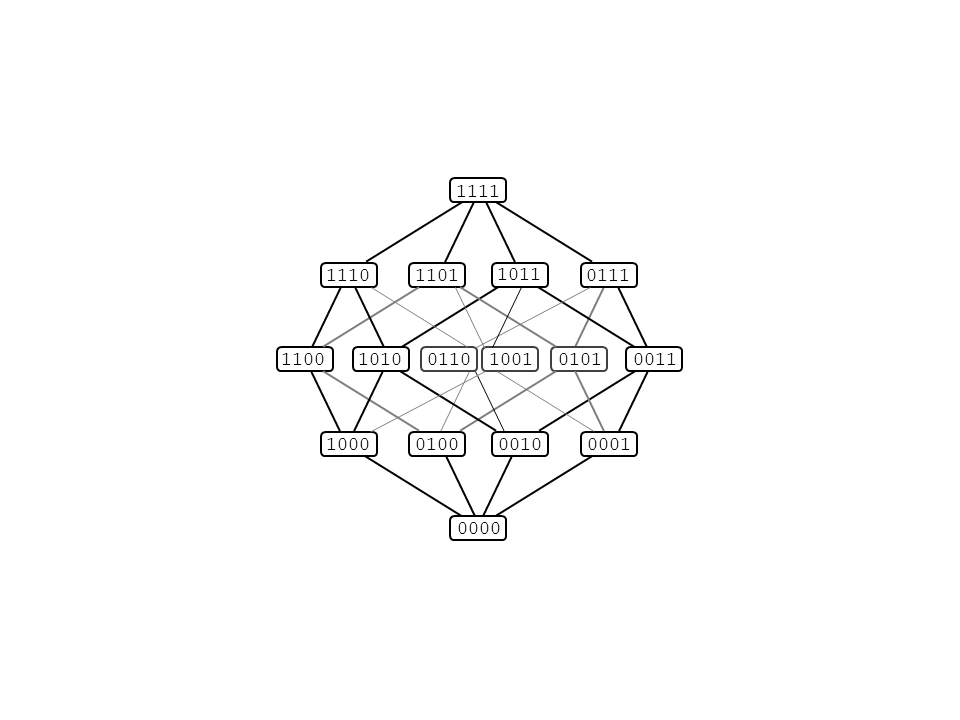}
%{C:/Users/NI/Documents/Doro/WriteUps/Graphics/4D_booleanLattice_binaryLabels}
\caption{A $4$-dimensional hypercube or Boolean lattice with its sixteen vertices labeled with $4$-bit codes (the vertices can also be labeled with their decimal equivalents or respective support sets). Boolean lattices are structures that represent \emph{both} genotypes and gene clusters, but are \emph{separate domains}.} \label{Fig:4-D_Boolean_cube}
\end{figure}

There is a rich algebraic background to this setup. We refer to O'Donnell~\cite{AnalysisBooleanFncs_ODonnell2014} for a full exposition. We hope to attract the attention of both biologists and mathematicians to this subject by quickly highlighting some topics. The set of binary strings of length $n$ can be represented geometrically as the vertices of a hypercube, with edges connecting two strings that dif\mbox{}fer in just one bit. The Boolean lattice for the case of $n=4$ bits is shown in Figure~\ref{Fig:4-D_Boolean_cube}.  The strings comprise a commutative group, called $\Int_2^n$, with an operation of pointwise addition of bits, modulo $2$.

Denote $\bin{i}_l$ as the binary state of the gene or factor positioned at locus $l$,\footnote{Position in the binary string does not correspond to position on the chromosome, and dif\mbox{}ferent positions may represent alleles from dif\mbox{}ferent chromosomes, or even alleles from mitochondrial alleles.} with $\bin{i}_l \in \set{\bin{0},\bin{1}} $. Then a \emph{genotype} can be represented by the $n$-bit string or code $\vect{\bin{i}_n\ldots\bin{i}_2\!\:\bin{i}_1}$.\footnote{The reverse ordering of the loci is not necessary, but conforms to base-$2$ positional notation which facilitates conversion to base-$10$.} This binary string has a \emph{decimal equivalent} $i = \sum_{l=1}^n \bin{i}_l\!\cdot\!2^{l-1}$, ranging from $0$ to $2^n-1$. Thus, a `$\bin{1}$' or `$\bin{0}$' at locus~$l$ respectively tells us whether or not the $2^{l-1}$ term is included in integer $i$. This motivates a third notation for strings that simply identifies the support set of loci containing a `$\bin{1}$'. Denote the \emph{set of integers} $\set{1,2,\ldots n}$ as $\vect{n}$. For the string associated with index $i$, its \emph{support set} is defined as $\mathcal{S}_i = \set{\,l\: |\: \bin{i}_l = \bin{1}} \subseteq \vect{n}$. For instance, when $n=4$ loci, the string $\vect{\bin{0101}}$ corresponds to decimal~$i=5$ and support set $\mathcal{S}_5 = \set{1,3}$. The product of two sets in this group is just their symmetric dif\mbox{}ference, which is analogous to the operation of bitwise addition.

These three dif\mbox{}ferent notations are used interchangeably to identify an individual organism's genotype, and permits representing its \emph{trait value}\footnote{If there are multiple organisms with the same genotype, then $t_i$ represents a mean of their trait values.} as either~$t_{\vect{\bin{i}_n\ldots\bin{i}_2\!\:\bin{i}_1}}$, $t_i$, or $t_{\mathcal{S}_i}$, depending on ease of exposition. For example, the \emph{full trait} $\bs{t}$ defined on the $n$-dimensional Boolean lattice is compactly expressed as $\bs{t}=\set{t_i}_{i=0}^{2^n-1}$ or $\bs{t} = \set{t_{\mathcal{S}_i}}_{\mathcal{S}_i\subseteq\vect{n}}$.\\

A primary goal of this paper is to discover interactions between genes/factors. Among the~$n$ loci of interest, there are $2^n$ ways to group them to study their interactions, e.g., as singletons, pairs, triples, etc. An abstract view of the ``clustering'' of $0\le k\le n$ genes is a \emph{$k$-tuple}, which can be represented by an $n$-bit string containing precisely $k$ `$\bin{1}$'s. To dif\mbox{}ferentiate these binary strings from the genotypes described above, we will reserve the index $j$ to refer to these ``cluster codes,'' and use index $i$ exclusively for genotypes. Hence, $\vect{\bin{j}_n\ldots\bin{j}_2\!\:\bin{j}_1}$ is a cluster code,\footnote{It is imperative to understand the subtle, but crucial dif\mbox{}ference between the binary strings representing genotypes and gene-cluster codes: in a genotype, $\bin{i}_l$ simply encodes the state of a binary allele/factor at locus $l$, whereas in a cluster code, $\bin{j}_l$ indicates whether or not the gene/factor at locus $l$ participates in a cluster.} where a `$\bin{1}$' or `$\bin{0}$' at $\bin{j}_l$ indicates \emph{whether or not the gene at locus~$l$ is a member of that cluster}, respectively. The $2^n$ dif\mbox{}ferent cluster codes can also be represented geometrically as vertices of a $n$-dimensional Boolean lattice\,---\,however, the Boolean lattice representing gene clusters is a dif\mbox{}ferent object than the one representing genotypes\,---\,they are fundamentally dif\mbox{}ferent domains.

As before, each cluster code has a decimal equivalent $j = \sum_{l=1}^n \bin{j}_l\!\cdot\!2^{l-1}$, and the participating loci can alternatively be represented by the support set $\mathcal{S}_j = \set{\,l\: |\: \bin{j}_l = \bin{1}} \subseteq \vect{n}$. For example, cluster code $\vect{\bin{1110}}$ corresponds to decimal~$j=14$ and support $\mathcal{S}_{14} = \set{2,3,4}$, which reveals that loci $l=2,3,4$ are members of this cluster. Cluster codes $\vect{\bin{j}_n\ldots\bin{j}_2\!\:\bin{j}_1}$, their equivalent decimal indices $j$ and support sets $\mathcal{S}_j$ are later used to index the Fourier transform and its coef\mbox{}ficients.

Gene clusters can be organized into so-called ``levels'' based on the cardinality of their support sets. For $0 \le k \le n$, define the \kth\ \emph{level} $\level{k}$ as the \emph{set of clusters whose support sets have cardinality $k$}:
\begin{equation} \label{Def:Level_Lk}
\level{k} \,=\, \BigSet{\; \vect{\bin{j}_n\ldots\bin{j}_2\!\:\bin{j}_1} \,\; \big| \,\; \abs{\mathcal{S}_j} = k\;}.
\end{equation}\\[-7pt]
In other words, level $\level{k}$ groups together all of the $k$-tuple clusters. Hence, level $\level{1}$ lists the codes for the \emph{individual} loci, level~$\level{2}$ contains the codes for \emph{pairs} of loci, and so on. It should be clear that the number of gene clusters in each level is ``$n$ choose $k$'': $\abs{\level{k}} = \textstyle\binom{n}{k}$. Returning to the example of $n=4$ loci, the vertices of the Boolean lattice in Figure~\ref{Fig:4-D_Boolean_cube} can also be viewed as the clusters of the corresponding gene network. Notice the $4$-bit codes are stratified into their five (ascending) levels~$\set{\level{k}}_{k=0}^4$, and that the cardinality of each level confirms $\abs{\level{k}} = \binom{4}{k}$.\\

\textbf{Definition:} A \emph{gene network} $\bs{g}$ is a real-valued function on a Boolean lattice of gene clusters.\\

Let the \emph{interaction exhibited by a gene cluster} be alternatively denoted by $g_{\vect{\bin{j}_n\ldots\bin{j}_2\!\:\bin{j}_1}}$, $g_j$, or $g_{\mathcal{S}_j}$. Like the trait $\bs{t}$, the \emph{full network of gene interactions} $\bs{g}$ can be concisely expressed as $\bs{g}=\set{g_j}_{j=0}^{2^n-1}$ or $\bs{g} = \set{g_{\mathcal{S}_j}}_{\mathcal{S}_j\subseteq\vect{n}}$. The small examples in Tables~\ref{Tab:2-loci_Dom_trait_gene_network},~\ref{Tab:2-loci_Inf_trait_gene_network},~\ref{Tab:Modular_trait_n=4}
(pages~\pageref{Tab:2-loci_Dom_trait_gene_network},~\pageref{Tab:2-loci_Inf_trait_gene_network},~\pageref{Tab:Modular_trait_n=4}, respectively) may help orient the reader to the various label notations and how traits and gene networks are functions of them (however, the support set $\mathcal{S}_i$ labels are omitted for traits since they are easiest understood in terms of their genotype~$\vect{\bin{i}_n\ldots\bin{i}_2\!\:\bin{i}_1}$).

\vspace{10pt}
Traits have a variety of aspects. As functions on the hypercube vertices, the trait variation in moving from one vertex to a neighbor, or more generally, in percolation through the entire lattice may be either smooth and gradual or rugged and varying. But, as functions on a basis of integers (coding a genotype in base-2), traits may be treated as column vectors. However, since there is a group multiplication on the basis, the set of traits is an algebra, a vector space endowed with a convolutional multiplication as well as vector addition.

When the range of a function is limited to two alternatives, traits have been extensively studied in the theoretical computer science literature under the name ``\emph{Boolean-valued Boolean functions}.'' This theoretical machinery can be readily generalized to ``\emph{real-valued Boolean functions}'' for quantitative traits~\cite{AnalysisBooleanFncs_ODonnell2014}.

Crucially, when a vector space has a basis which is a commutative group, there will be a Fourier transform, which rewrites the original in a new basis. In our case, the new basis consists of homomorphisms of the group of strings to the set $\set{+1,-1}$, and ``interesting combinatorial properties of a Boolean function can be `read of\mbox{}f' from its Fourier coef\mbox{}ficients''~\cite[p.~$26$]{AnalysisBooleanFncs_ODonnell2014}.

% --------------------------------------------------------------
% ----------------------------------------------------------------------------------
\subsection{The Sylvester-Hadamard matrix and Fourier transform} \label{Sec:HadMatrix_Fourier}
% ----------------------------------------------------------------------------------
% --------------------------------------------------------------
Let us construct the Fourier transform of trait space in concrete terms. The \emph{Sylvester-type\footnote{
These matrices were introduced as "tessellated pavements" by J.J.~Sylvester in $1867$, who commends their versatility, ``furnishing interesting food for thought, or a substitute for the want of it, alike to the analyst at his desk and the fine lady in her boudoir''~\cite{ThoughtsInverseOrthogMatrices_Sylvester1867}. They were generalized to non-powers-of-two by Hadamard in $1893$, and independently proposed as \emph{continuous functions} by Walsh in $1923$; see~\cite{WalshFourierStatApps_Stoffer1991} for an excellent historical review. Note, power-of-two Hadamard matrices and their respective transforms are often referred to \emph{en masse} simply as ``Walsh-Hadamard''  matrices and transforms, although formally, this connotes a dif\mbox{}ferent ordering of the rows and columns from the Sylvester-type defined in~(\ref{Def:Syl_Had_matrix}).} Hadamard matrix}~$\bs{\Syl}_n$ of order~$2^n$ is defined recursively for $n\ge1$ by
\begin{equation} \label{Def:Syl_Had_matrix}
   \bs{\Syl}_n \,=\;
    \left[
         \begin{array}{rr}
           \bs{\Syl}_{n-1} & \bs{\Syl}_{n-1} \\[5pt]
           \bs{\Syl}_{n-1} & -\bs{\Syl}_{n-1} \\
         \end{array}
       \right]
\end{equation}
where $\bs{\Syl}_0 = 1$. Row and column indices should be labeled $0$ to $2^n-1$, or in their binary equivalents. By definition, all Hadamard matrices consist of $\pm1$ entries and are orthogonal~\cite[p.~$9$]{Horadam_HadamardMatrices}, thus
\begin{equation} \label{Eqn:Syl_Orth_Relation}
\bs{\Syl}_n\bs{\Syl}_n^\top \,=\, 2^n \bs{I}
\end{equation}
where the superscript ``$\top$'' denotes matrix transposition, and $\bs{I}$ is the identity matrix of order $2^n$. Further, it is well known that Sylvester-type matrices are symmetric: $\bs{\Syl}_n = \bs{\Syl}_n^\top$, hence
\begin{equation} \label{Eqn:Syl_Inv_Matrix}
\bs{\Syl}_n^{-1} \,=\, \bs{\Syl}_n/2^n.
\end{equation}
More details on Hadamard matrices and their use in group theory can be found in~\cite{Horadam_HadamardMatrices}.
Henceforth, except when necessary, we will let  $\bs{\Syl} = \bs{\Syl}_n$, dropping the subscript $n$ for ease of exposition.\\

We define the \emph{gene network}~$\bs{g}$ associated with a trait~$\bs{t}$ to be its (forward) Fourier transform $\mathcal{F}\gr{\bs{t}}$. This can be formally expressed as the matrix-vector multiplication
\begin{equation} \label{Eqn:g=Ht}
\bs{g} \,=\, \mathcal{F}\gr{\bs{t}} \,=\, \bs{\Syl} \bs{t}/2^n.
\end{equation}
The coef\mbox{}ficients of the vector $\bs{g}$ are \emph{a spectrum of interactions} into which the trait $\bs{t}$ is broken. The $2^n$ factor in the denominator of~(\ref{Eqn:g=Ht}) permits each entry of $\bs{g}$ to be viewed as a unique weighted average of the trait being examined. That is, each of the $2^n$ spectral components are associated with a particular row of the Hadamard matrix, which specifies a pattern of signs\,---\,these are the weights (i.e., signed factors of~$1/2^n$) used in each average over the trait.

The choice of the labels `$\bin{0}$' or `$\bin{1}$' at a particular locus is a matter of convenience. It is easy to show that flipping the bit of an arbitrary locus across all genotypes will only change the sign, and not the magnitude, of its associated gene network coef\mbox{}ficients. Given a gene network $\bs{g}$, we can take its inverse Fourier transform using (\ref{Eqn:Syl_Inv_Matrix}) to find its full trait for the whole population of genotypes:
\begin{equation} \label{Eqn:t=Hg}
\bs{t} \,=\, \mathcal{F}^{-1}\gr{\bs{g}} \,=\, \bs{\Syl}\bs{g}.
\end{equation}
This shows that the trait value for a particular genotype is reconstructed as the appropriate weighted (i.e., the pattern of $\pm1$'s in the associated row of $\bs{\Syl}$) combination of the gene interactions.\\

Let us now illustrate the connection between: (i) the three notations for the indices of $\bs{g}$, (ii) the binary string notation of $\bs{t}$, and (iii) the sign patterns of matrix $\bs{\Syl}$. For any $n\ge1$, we have for $\bs{\Syl}$ (omitting the obvious unity factors) that the top $j=0$ row is $\vect{+, +, +, +, \ldots}$, row $j=1$ is $\vect{+, -, +, -, \ldots}$, row $j=2$ is $\vect{+, +, -, -, \ldots}$, row $j=3$ is $\vect{+, -, -, +, \ldots}$, and so on. Let the symbol `$\bin{*}$' serve as a wildcard for a `$\bin{0}$' or `$\bin{1}$' in the loci that we wish to ``ignore'' across the genomes. Then summing over all possibilities for the wildcards (i.e., over all genotypes), we can list the first four elements of the spectrum:
%\begin{equation*}
%\begin{array}{llllcll}
%g_0 & = & g_{\vect{\bin{0\ldots000}}} & = & g_\varnothing & = & \Gr{\Sigma\,t_{\vect{\bin{*\ldots***}}}}/2^n \\[4pt]
%g_1 & = & g_{\vect{\bin{0\ldots001}}} & = & g_{\set{1}} & = &  \Gr{\Sigma\,t_{\vect{\bin{*\ldots**0}}} - t_{\vect{\bin{*\ldots**1}}}}/2^n \\[4pt]
%g_2 & = & g_{\vect{\bin{0\ldots010}}} & = & g_{\set{2}} & = &  \Gr{\Sigma\,t_{\vect{\bin{*\ldots*0*}}} - t_{\vect{\bin{*\ldots*1*}}}}/2^n \\[4pt]
%g_3 & = & g_{\vect{\bin{0\ldots011}}} & = & g_{\set{1,2}} & = &  \Gr{\Sigma\,t_{\vect{\bin{*\ldots*00}}} - t_{\vect{\bin{*\ldots*01}}} - t_{\vect{\bin{*\ldots*10}}} + t_{\vect{\bin{*\ldots*11}}}}/2^n \\[2pt]
%\end{array}
%\end{equation*}
\begin{equation*}
\begin{array}{llllcll}
g_0 & = & g_{[\texttt{0\ldots000}]} & = & g_\varnothing & = & \big(\Sigma \, t_{[\texttt{*\ldots***}]}\big)/{2^n} \\[4pt]
g_1 & = & g_{[\texttt{0\ldots001}]} & = & g_{\{1\}} & = & \big(\Sigma \, t_{[\texttt{*\ldots**0}]} - t_{[\texttt{*\ldots**1}]}\big)/{2^n} \\[4pt]
g_2 & = & g_{[\texttt{0\ldots010}]} & = & g_{\{2\}} & = & \big(\Sigma \, t_{[\texttt{*\ldots*0*}]} - t_{[\texttt{*\ldots*1*}]}\big)/{2^n} \\[4pt]
g_3 & = & g_{[\texttt{0\ldots011}]} & = & g_{\{1,2\}} & = & \big(\Sigma \, t_{[\texttt{*\ldots*00}]} - t_{[\texttt{*\ldots*01}]} - t_{[\texttt{*\ldots*10}]} + t_{[\texttt{*\ldots*11}]}\big)/{2^n} \\[2pt]
\end{array}
\end{equation*}
%\begin{equation*}
%\begin{array}{llllcll}
%g_0 & = & g_{[\texttt{0\ldots000}]} & = & g_\varnothing & = & \textstyle{\frac{1}{2^n}} \Sigma \, t_{[\texttt{*\ldots***}]} \\[4pt]
%g_1 & = & g_{[\texttt{0\ldots001}]} & = & g_{\{1\}} & = & \textstyle{\frac{1}{2^n}} \Sigma \, t_{[\texttt{*\ldots**0}]} - t_{[\texttt{*\ldots**1}]} \\[4pt]
%g_2 & = & g_{[\texttt{0\ldots010}]} & = & g_{\{2\}} & = & \textstyle{\frac{1}{2^n}} \Sigma \, t_{[\texttt{*\ldots*0*}]} - t_{[\texttt{*\ldots*1*}]} \\[4pt]
%g_3 & = & g_{[\texttt{0\ldots011}]} & = & g_{\{1,2\}} & = & \textstyle{\frac{1}{2^n}} \Sigma \, t_{[\texttt{*\ldots*00}]} - t_{[\texttt{*\ldots*01}]} - t_{[\texttt{*\ldots*10}]} + t_{[\texttt{*\ldots*11}]} \\[2pt]
%\end{array}
%\end{equation*}
Notice, \emph{the locations of the `$\bin{1}$'s in each cluster code act as binary flags indicating which loci of the genomes are to be analyzed}. Conversely, the locations of the `$\bin{0}$'s in the cluster codes mean that these loci of the genomes are to be ignored, regarded as fixed in the background. For example, consider row $j=1$, which corresponds to locus $l=1$, and observe that the positive signs in the sequence occur when there is a `$\bin{0}$' allele in locus $l=1$ of the genotypes; conversely, negative signs in this sequence occur when there is a `$\bin{1}$' in locus $l=1$. Similarly, row $j=2$ assigns positive/negative signs to precisely those genotypes with a `$\bin{0}$'/`$\bin{1}$' allele in locus $l=2$, and so on. Thus, we see how an arbitrary cluster code is directly related to \emph{the ``untangling'' property of the Fourier transform:} its associated sign pattern yields the relevant weighted average that ``teases'' apart (the complex and interrelated relationships between) the trait values into the isolated ef\mbox{}fect due to that unique cluster.

What information can be gleaned from these values of $\bs{g}$? Employing the support set notation, the $g_\varnothing$ coef\mbox{}ficient is just the arithmetic mean of the trait, owing to its associated row of entirely $+1$ weights. Coef\mbox{}ficient $g_{\set{1}}$ measures the dif\mbox{}ference between the average of the $2^{n-1}$ trait values whose genotypes have a `$\bin{0}$' at locus $l=1$ and average of the $2^{n-1}$ trait values with a `$\bin{1}$' at locus $l=1$. This is a measure of the direct ef\mbox{}fect of locus $l=1$. Similarly, $g_{\set{2}}$ measures the individual ef\mbox{}fect from the gene at locus $l=2$, and, in general, $g_{\set{l}}$ gives the direct allele ef\mbox{}fect from locus~$l$.

An ambiguity now arises for the cluster with two loci. We can picture $g_{\set{1,2}}$ as an interactive ef\mbox{}fect of varying locus $l=2$ on the already established ef\mbox{}fect of locus $l=1$, by writing
%$$
%g_{\set{1,2}} \;=\; \Gr{\Sigma\,\gr{t_{\vect{\bin{*\ldots*00}}} - t_{\vect{\bin{*\ldots*01}}}} - \gr{t_{\vect{\bin{*\ldots*10}}} - t_{\vect{\bin{*\ldots*11}}}}}/2^n.
%$$
$$
g_{\{1,2\}} \;=\; \big(\Sigma \, (t_{[\texttt{*\ldots*00}]} - t_{[\texttt{*\ldots*01}]}) - (t_{[\texttt{*\ldots*10}]} - t_{[\texttt{*\ldots*11}]})\big)/2^n.
$$
%$$
%g_{\{1,2\}} \;=\; \textstyle{\frac{1}{2^n}}\Sigma \, (t_{[\texttt{*\ldots*00}]} - t_{[\texttt{*\ldots*01}]}) - (t_{[\texttt{*\ldots*10}]} - t_{[\texttt{*\ldots*11}]}).
%$$
However, this is patently the same as the ef\mbox{}fect of locus $l=1$ on locus $l=2$:
%$$
%g_{\set{1,2}} \;=\; \Gr{\Sigma\,\gr{t_{\vect{\bin{*\ldots*00}}} - t_{\vect{\bin{*\ldots*10}}}} - \gr{t_{\vect{\bin{*\ldots*01}}} - t_{\vect{\bin{*\ldots*11}}}}}/2^n.
%$$
$$
g_{\{1,2\}} \;=\; \big(\Sigma \, (t_{[\texttt{*\ldots*00}]} - t_{[\texttt{*\ldots*10}]}) - (t_{[\texttt{*\ldots*01}]} - t_{[\texttt{*\ldots*11}]})\big)/2^n.
$$
%$$
%g_{\{1,2\}} \;=\; \textstyle{\frac{1}{2^n}}\Sigma \, (t_{[\texttt{*\ldots*00}]} - t_{[\texttt{*\ldots*10}]}) - (t_{[\texttt{*\ldots*01}]} - t_{[\texttt{*\ldots*11}]}).
%$$
The action of locus $l=1$ on locus $l=2$ also equals the action of locus $l=2$ on locus $l=1$. Action equals reaction! For general cluster pairs, an even-handed description simply says that $g_{\set{l,l'}}$ represents the interaction between loci~$l$ and~$l'$, adding all trait values whose subscript has an even number of `$\bin{1}$' entries within bits~$\bin{i}_l$ and~$\bin{i}_{l'}$ and subtracting all those with an odd number of `$\bin{1}$' entries. This measures a non-linearity in the ef\mbox{}fects between loci~$l$ and~$l'$. If~$l$ and~$l'$ are homologous loci and `$\bin{1}$' represents a dominant allele, then $g_{\set{l,l'}}$  will be a number which quantifies a saturation interaction between loci~$l$ and~$l'$; this is illustrated in Table~\ref{Tab:2-loci_Dom_trait_gene_network} of Example~\ref{Ex:2-loci_Dom_trait} on page~\pageref{Tab:2-loci_Dom_trait_gene_network}.

In general, for some subset of loci $\mathcal{S}_j$, the coef\mbox{}ficient $g_{\mathcal{S}_j}$  sums the trait over entries with an even number of `$\bin{1}$' alleles, minus those with an odd number of such entries, and we shall consider $g_{\mathcal{S}_j}$ as a quantitative measurement of non-linearities attributable to the interaction of loci within the set $\mathcal{S}_j$.

We can see that, as promised, the Fourier coef\mbox{}ficients do indeed express important features of the trait. They are arranged in a hierarchy, where the level of a coef\mbox{}ficient $g_{\mathcal{S}_j}$ is just the cardinality of $\mathcal{S}_j$, as described earlier. Of course there are multiple plausible ways to express such multi-level epistasis, with various choices of sign and of scaling by powers-of-two, as explored in the delightful papers~\cite{ContextDependenceMutations_Poelwijk2016, LearningPatternEpistasis_Poelwijk2019}, but our chosen scheme has several elegant features, which permit compressive sensing.

The Fourier transform is self-inverse, isometric, and maximally incoherent.
\begin{itemize}
\item The Sylvester-Hadamard matrix is, up to rescaling, its own inverse (see (\ref{Eqn:Syl_Orth_Relation})). The trait vector $\bs{t}$ can be reconstructed from the gene network $\bs{g}$ by a second application of the Hadamard matrix. We previously mentioned the rescaling factor $1/2^n$ as the uniform averaging constant applied to interpret entries of $\bs{g}$, but it may be treated as a choice of units in network space. All that is of importance is the relative size of the coef\mbox{}ficients of $\bs{g}$, not absolute size.

\item The determinant of $\bs{\Syl}$ is $2^n$, and all its rows and columns  are orthogonal. Therefore, the maps from trait space to gene network space and back are, after rescaling, isometries, preserving length and angle of vectors by Parseval's and Plancherel's theorems. Thus, measurement errors and approximations that are small in one space, remain small in the other, using the $\ell_2$-norm (root mean squared sums). This assures us that very small entries in the network space may be set to zero with small ef\mbox{}fect on the corresponding trait. If a trait is concentrated in a small number of network coef\mbox{}ficients, then those few large coef\mbox{}ficients can be used to generate an excellent approximation of the original trait.  This would not necessarily work for an arbitrary invertible transform, where small ef\mbox{}fects in one space could have large ef\mbox{}fects in the other.

\item The Hadamard matrix as a sensing modality is maximally incoherent with respect to the standard basis, i.e., the identity matrix. Since the gene network is sparse relative to the standard basis, maximal incoherence means minimal observations in the trait space yield all the information contained in the gene network space. See Section~{\ref{Sec:CS_Implications_traits}} for more details.
\end{itemize}

% ==============================================================
% ==================================================================================
\section{Comparison with other networks}
% ==================================================================================
% ==============================================================
As the number of genes af\mbox{}fecting a complex trait increases, it becomes natural to visualize gene interactions as a graph, and to seek general features of their architecture. The topology of this paper's gene network can be visualized as a weighted simplicial complex. This means that each gene locus~$l$ can be pictured as a point, weighted by its corresponding level~$\level{1}$ Fourier coef\mbox{}ficient, each level~$\level{2}$ interaction by an edge between the two loci involved, say, $l$ and $l'$, with the weight $g_{\set{l,l'}}$. A level~$\level{3}$ interaction between loci~$l$, $l'$ and $l''$ is pictured as a triangle, weighted by $g_{\set{l,l',l''}}$. Higher order interactions can be represented by higher dimensional simplices, as desired.

Our gene network quantifies non-linear relationships. It does resemble some other better known gene networks so that common structural features can be expected~\cite[Ch.~13]{StructureComplexNetworks_Estrada2011}.

% --------------------------------------------------------------
% ----------------------------------------------------------------------------------
\subsection{Coexpression networks and modules} \label{Sec:Compare:CoexpressionNetsModules}
% ----------------------------------------------------------------------------------
% --------------------------------------------------------------
A gene coexpression network represents genes by points, with an undirected edge for genes with similar activity profiles. Such networks demonstrate clustering of genes~\cite{HumanGeneCoexpressionLandscape_Prieto2008} and it is natural to expect that these clusters reflect an organization of genes into dif\mbox{}ferent functional modules. Gene interactions whose support sets span two or more distinct modules should be rare. Fourier coef\mbox{}ficients will have large cancellation ef\mbox{}fects when an irrelevant locus is included in the support because of the varying parity of relevant loci.

There is considerable evidence that biological functions are typically organized into modules, each with an associated suite of genes~\cite{MolecularModularCellBio_Hartwell1999}, breaking a task into an array of subtasks. In turn, subtasks may themselves be composite, leading to a hierarchical structure. This organizing principle facilitates evolution~\cite{ModularityNecEvolvability_Hansen2003}, since a submodule may be modified without inducing global complications. This explains why those exceptional proteins which interact with many other proteins are very stable throughout evolution~\cite{PleiotropyPreservationPerfection_Waxman1998}.

It is no trivial matter even to detect modules in complex networks~\cite{PerformanceModularityMax_Good2010}, and Kleinberg's theorem rules out an algorithm with all three desirable features of scale invariance, richness and consistency~\cite{ImpossThmClustering_Kleinberg2003}. Modified ``spectral redemption'' techniques~\cite{SpectralRedemption_Krzakala2013} may be applied to detect modules in very simple gene networks.

% --------------------------------------------------------------
% ----------------------------------------------------------------------------------
\subsection{Regulatory networks and sparsity}
% ----------------------------------------------------------------------------------
% --------------------------------------------------------------
A second type of gene network is the gene regulatory network. Again, a node represents a gene locus, but there is a directed edge from each regulatory gene to each target. Judea Pearl points out~\cite[p.~$75$]{Why_Pearl2018} that similar causal networks were first used by Sewall Wright in $1920$~\cite{RelImportHeredityEnv_Wright1920}. These networks have a hub and spoke structure, with regulatory genes each surrounded by a cluster of target genes, with a fat-tailed distribution of outdegrees~\cite{StructureComplexNetworks_Estrada2011}. These network of directed edges dif\mbox{}fer from the undirected edges of this paper, but an edge in a gene regulatory network is very likely a edge in our gene network, since it represents a non-linear interaction. Our gene network is specific to some trait and makes no distinction between regulatory and target genes.

The ``hub-and-spoke'' structure of gene regulatory networks resembles that of Barab\'asi-Albert preferential attachment graphs~\cite{StatMechComplexNetwork_Barabasi2002}, which have a scale-free degree distribution. Such graphs evolve by inserting new nodes of fixed valency. The new nodes prefer attachment to existing nodes of high valency. This leads to scale-free degree distribution. Their evolution parallels the evolution of gene regulatory networks by duplicating a node and then specializing its function by reassigning edges. Preferential attachment graphs are sparse, since there is a fixed ratio of edges to nodes. Leclerc~\cite{SurvivalSparsestParsimonious_Leclerc2008} summarizes data on eight gene regulatory networks showing that such networks are both sparse and robust, resisting perturbation from either from environmental or mutational changes. These results accord well with our prior observations that low-level concentration of Fourier coef\mbox{}ficients promotes robust traits.

% ==============================================================
% ==================================================================================
\section{Low-level concentration and roughness}
% ==================================================================================
% ==============================================================
There are some quite reasonable assumptions about the general nature of a realistic gene network. One can expect a great deal of information about a trait to be conveyed by its average over all genotypes (its level $\level{0}$ network coef\mbox{}ficient), or by the average ef\mbox{}fect of one allele versus another at some locus (level $\level{1}$ ef\mbox{}fects). Similarly, the epistatic ef\mbox{}fects of a few loci are expected, but it becomes increasingly hard to imagine a mechanism by which the parity of a large set of marked alleles makes much dif\mbox{}ference on average. Therefore, it is natural to expect that the transform of a trait into network space compresses the data into lower levels. Of course, sparsity or compressibility is an obvious consequence of low-level concentration. Evolvability and modularity are two features of a trait that naturally usher in low-level concentration. We discuss evolvability below, and the ef\mbox{}fects of modularity in Section~\ref{Sec:Modular_traits}.

The debate between gradual and abrupt change is central to evolutionary thinking~\cite{darwin1859}. Roughness has been much explored for the terrain of the trait ``fitness''~\cite{MeasuringRuggedness_VanCleve2015}. Evolution studies a population distributed throughout a Boolean lattice of genomes, gradually dif\mbox{}fusing toward a fitness optimum. A \emph{rugged} fitness terrain is one where fitness makes big jumps with small mutations, while fitness changes gradually on a \emph{smooth} terrain.  Populations evolving on a smooth terrain can evolve towards an optimum more directly, while more exploration is required for rougher terrains~\cite{RuggedLandscapes_Kauffman1987, RolesMutationInbreeding_Wright1932}.

% --------------------------------------------------------------
% ----------------------------------------------------------------------------------
\subsection{Local, level, and total roughness}
% ----------------------------------------------------------------------------------
% --------------------------------------------------------------
Every genome can be associated with a subset of loci (i.e., the support set of loci containing a `$\bin{1}$'), and an edge is associated with two sets which dif\mbox{}fer only in a singleton, as depicted in Figure~\ref{Fig:4-D_Boolean_cube}. An edge is the smallest possible mutation\,---\,only one locus mutates. So, if $\mathcal{S}_i$ and $\mathcal{S}_{i'}$ share an edge, the symmetric dif\mbox{}ference $\mathcal{S}_i\!\bigtriangleup\!\mathcal{S}_{i'}$ will be a singleton. This edge contributes $\Gr{\gr{t_{\mathcal{S}_i}-t_{\mathcal{S}_{i'}}}/2}^2$ to the roughness of the trait~$\bs{t}$, as discussed in~\cite{AnalysisBooleanFncs_ODonnell2014, NoiseSensBoolFncsPerc_Garban2014}. Of course, some locus~$l$ may contribute little to roughness, if the average ef\mbox{}fect of varying alleles at locus~$l$ changes the trait by little. We define the \emph{local roughness of trait~$\bs{t}$ at locus~$l$} by
\begin{equation} \label{LOCAL_roughness_trait}
\inflLOC{l}{\bs{t}} \;= \sum_{\substack{\mathcal{S}_i\bigtriangleup\mathcal{S}_{i'}=\set{l}\\ \mathcal{S}_i,\mathcal{S}_{i'}\subseteq\vect{n}}}
\AutoGr{\frac{t_{\mathcal{S}_i}-t_{\mathcal{S}_{i'}}}{2}}^2.
\end{equation}
Each edge is counted twice in this formulation, because we can switch $\mathcal{S}_i$ and $\mathcal{S}_{i'}$. The \emph{full trait roughness} is the sum of all local roughnesses:
\begin{equation} \label{TOTAL_roughness_trait}
\inflTOT{\bs{t}} \;=\; \sum_{l=1}^n R_l\gr{\bs{t}} \;= \sum_{\substack{\abs{\mathcal{S}_i\bigtriangleup\mathcal{S}_{i'}}=1\\ \mathcal{S}_i,\mathcal{S}_{i'}\subseteq\vect{n}}} \AutoGr{\frac{t_{\mathcal{S}_i}-t_{\mathcal{S}_{i'}}}{2}}^2.
\end{equation}

Roughness can be considered a form of ``energy'' on account of the squared terms in~(\ref{LOCAL_roughness_trait}). Local roughness at locus~$l$ has been called~\cite{AnalysisBooleanFncs_ODonnell2014, NoiseSensBoolFncsPerc_Garban2014} the ``influence'' of~$l$, while ``energy'' has a number of aliases such as ``average sensitivity,'' ``total influence,'' ``normalized edge boundary,'' and ``responsiveness''~\cite[p.~$7$]{TopicsBooleanFns_ODonnell2008}. In a random walk through the hypercube, the square root of roughness estimates the average (specifically, the root-mean-square) change at each step.

One of our goals is to evaluate the roughness of a trait's landscape \emph{directly from its gene network}. There is a simple expression of local roughness~(\ref{LOCAL_roughness_trait}) in terms of $\bs{g}$, the Fourier transform  of our trait~$\bs{t}$. It is based on Parseval's theorem and the expression of the dif\mbox{}ference operator as a convolution operator~\cite{AnalysisBooleanFncs_ODonnell2014, NoiseSensBoolFncsPerc_Garban2014}. Thus, the \emph{local influence of locus~$l$} on roughness can be expressed in the Fourier domain as
\begin{equation} \label{Eqn:LOCAL_roughness_Fourier}
\inflLOC{l}{\bs{t}} \;=  \sum_{\substack{l\in\mathcal{S}_j\\ \mathcal{S}_j\subseteq\vect{n}}} g_{\mathcal{S}_j}^2, \qquad l\in\vect{n}.
\end{equation}
In words, the influence of locus~$l$ is the accumulation of interaction energy from all of the gene clusters of which it is a member. Similarly, the \emph{influence of level~$\level{k}$} (see~(\ref{Def:Level_Lk})) on the roughness of a trait's landscape is
\begin{equation} \label{Eqn:LEVEL_roughness_Fourier}
\inflLEV{k}{\bs{t}} \;= \sum_{\substack{\abs{\mathcal{S}_j}=k\\ \mathcal{S}_j\subseteq\vect{n}}} k\cdot g_{\mathcal{S}_j}^2, \qquad 0 \le k \le n
\end{equation}
where the factor of~$k$ occurs because each $k$-tuple cluster has a $k$-fold presence (i.e., $k$ distinct local influences).
Thus, level index~$k$ serves as a sort of ``moment-arm'' giving more weight to higher-level interactions.
Low level concentration leads to local smoothness because the terms in~(\ref{Eqn:LEVEL_roughness_Fourier}) with a large multiplier $k$ are absent or suppressed. Also note that the sole level~$\level{0}$ coef\mbox{}ficient $g_\varnothing$ just measures the average ``height'' of the trait landscape and so it should not exert any influence on its roughness. This is indeed the case: we will always have $\inflLEV{0}{\bs{t}} = 0$ due to its $k=0$ multiplier. Nonetheless, for the sake of completeness, we include the~$\level{0}$ case in~(\ref{Eqn:LEVEL_roughness_Fourier}) to show the influence over all levels.

It follows immediately that the \emph{total landscape roughness for a trait} is simply the aggregation of all local influences, or of all level influences:
\begin{equation} \label{Eqn:TOTAL_roughness_Fourier}
\inflTOT{\bs{t}} \;=\; \sum_{l=1}^n \inflLOC{l}{\bs{t}} \;=\; \sum_{k=0}^n \inflLEV{k}{\bs{t}} \;=\; \sum_{\mathcal{S}_j\subseteq\vect{n}} \abs{\mathcal{S}_j}\cdot g_{\mathcal{S}_j}^2.
\end{equation}
Therefore, Fourier components of higher level make an outsize contribution to a trait's total roughness.\footnote{We can even go so far as to say that the mid-to-high levels have \emph{an unfair advantage} to influence the total roughness~$\inflTOT{\bs{t}}$ due to the multiplier~$\abs{\mathcal{S}_j}$ in the right-hand side of~(\ref{Eqn:TOTAL_roughness_Fourier}). That is, a trait can only have a total roughness~$\inflTOT{\bs{t}}$ that is relatively small if the lower levels dominate, and the mid-high levels have, at most, small contribution.} Equivalently, a trait that varies smoothly should have its Fourier coef\mbox{}ficients concentrated in lower levels. One expects relatively few very rough loci, which will be highly conserved, while most loci should be smooth and hence more capable of evolution.

% --------------------------------------------------------------
% ----------------------------------------------------------------------------------
\subsection{Relative roughnesses}
% ----------------------------------------------------------------------------------
% --------------------------------------------------------------
We need a rubric to assess where a trait's roughness~$\inflTOT{\bs{t}}$ falls within the \emph{smooth--rugged continuum}. This is best determined by comparing it to the energy contained in its variance~$\var{\bs{t}}$. It can be shown~\cite{AnalysisBooleanFncs_ODonnell2014} that a trait's variance is related to its Fourier transform by
\begin{equation} \label{Eqn:var_Fourier}
\var{\bs{t}} \,= \sum_{\mathcal{S}_j\neq\varnothing} g_{\mathcal{S}_j}^2.
\end{equation}
The contribution of coef\mbox{}ficient $g_\varnothing$ is excluded as it represents substraction of the arithmetic mean in the traditional formulation of variance.

Now we can define the \emph{relative local and level influences} on roughness, respectively, as
\begin{equation} \label{Eqn:REL_LOCAL_LEVEL_roughness}
\RELinflLOC{l}{\bs{t}} \,=\, \frac{\inflLOC{l}{\bs{t}}}{\var{\bs{t}}}, \qquad
\RELinflLEV{k}{\bs{t}} \,=\, \frac{\inflLEV{k}{\bs{t}}}{\var{\bs{t}}}
\end{equation}
which yields the \emph{relative total roughness}
\begin{equation} \label{Eqn:REL_TOTAL_roughness}
\RELinflTOT{\bs{t}} \;=\; \sum_{l=1}^n \RELinflLOC{l}{\bs{t}} \;=\;  \sum_{k=0}^n \RELinflLEV{k}{\bs{t}} \;=\; \frac{\inflTOT{\bs{t}}}{\var{\bs{t}}} \;=\; \frac{\sum_{\mathcal{S}_j\subseteq\vect{n}} \!\:\abs{\mathcal{S}_j}\cdot g_{\mathcal{S}_j}^2}{\sum_{\mathcal{S}_j\neq\varnothing} \!\:g_{\mathcal{S}_j}^2}.
\end{equation}
Thus the relative total roughness is a weighted and normalized energy, which falls in the range of
\begin{equation} \label{Eqn:REL_TOTAL_roughness_LIMITS}
1 \;\le\; \RELinflTOT{\bs{t}} \;\le\; n.
\end{equation}
In fact, the right-hand side of~(\ref{Eqn:REL_TOTAL_roughness}) is in the form of a ``center of mass'' and so $\RELinflTOT{\bs{t}}$ can be interpreted as the ``ef\mbox{}fective level'' with the most influence, i.e., where the energy of the gene network is ef\mbox{}fectively concentrated. Clearly, the lower and upper limits of~(\ref{Eqn:REL_TOTAL_roughness_LIMITS}) occur when all of the gene network's energy (ignoring~$\level{0}$) is either concentrated in level~$\level{1}$ or~$\level{n}$, respectively. In this sense, we can claim that a trait whose \emph{$\RELinflTOT{\bs{t}}$ is closer to~$1$ has a very smooth landscape}, while it is \emph{quite rugged if $\RELinflTOT{\bs{t}}$ is closer to~$n$}. In summary, smooth traits are likely to occur near a stable, local evolutionary maximum, which are more insensitive to ambient fluctuations, while more rugged fitness landscapes promote greater diversity among a population~\cite{SurvivalSparsestParsimonious_Leclerc2008}.

Whether absolute or relative, the local and level influences and the total roughness each provide dif\mbox{}ferent insights into a trait via its fitness landscape. Obviously, the local influence is the most granular and the total roughness is the most global; in that sense, the level influence can be seen as a mediator between the two extremes.

We next provide some examples demonstrating these tools to appraise a trait's roughness. Examples~\ref{Ex:2-loci_Dom_trait} and~\ref{Ex:2-loci_Inf_trait} are extremely small, $2$-loci traits that illustrate the ideas in a simple and straightforward manner. Later, in Example~\ref{Ex:4-loci_ModularTrait} of Section~\ref{Sec:4-loci_Modular_example}, we analyze a $4$-loci trait that has a modular structure. Example~\ref{Ex:Entacmaea_trait} in Section~\ref{Sec:Entacmaea_example} is a real-world trait with $13$ loci. For the larger examples, we will see that smooth traits coincide with low-level concentration, and are therefore sparse.

% --------------------------------------------------------------
% ----------------------------------------------------------------------------------
\subsection{Very small $2$-loci examples}
% ----------------------------------------------------------------------------------
% --------------------------------------------------------------
Fitness landscapes can be visualized as the graph of a function on the hypercube. The  Fourier coef\mbox{}ficients can be used as the coef\mbox{}ficients of a polynomial that interpolates the trait values at hypercube vertices~\cite{AnalysisBooleanFncs_ODonnell2014}. Let us work two examples, for well-known trait landscapes, each involving two loci.

\begin{example} \label{Ex:2-loci_Dom_trait}
The first example is the classic picture of a dominant trait $\bs{t}_\text{dom}$. Without loss of generality, such a trait can be considered a Boolean-valued function, recorded in the left side of Table~\ref{Tab:2-loci_Dom_trait_gene_network}, with the \ith\ trait value $t_i$ as a function of the \ith\ genotype. Notice how the dominance of the trait is expressed so long as at least either $\bin{i}_1$ or $\bin{i}_2$ has a `$\bin{1}$'. The fitness landscape of this trait is seen in Figure~\ref{Fig:Dom_trait}. Although this example is trivially small, it is apparent that the landscape lacks variability: it looks fairly smooth. The following analysis confirms this.

First we need the associated gene network. From~(\ref{Eqn:g=Ht}) and~(\ref{Def:Syl_Had_matrix}) with $n=2$, we have $\bs{g}_\text{dom} = \bs{\Syl}_2\bs{t}_\text{dom}/2^2$, shown in the right side of Table~\ref{Tab:2-loci_Dom_trait_gene_network}, where the \jth\ interaction $g_j$ is a function of the \jth\ gene cluster. Utilizing support set notation, we notice $g_{\set{1}} = g_{\set{2}} = g_{\set{1,2}}$, which is an example of nonlinear saturation, i.e., the ef\mbox{}fect when both loci participate, seen in $g_{\set{1,2}}$, is the same as when just one loci participates, seen in $g_{\set{1}}$ and $g_{\set{2}}$.\\
\begin{table}[!h]
\centering
\begin{tabular}{|c|c||c|}
    \multicolumn{3}{c}{{Trait} $\bs{t}_\text{dom}$} \\
    \hline
    \multicolumn{1}{|c}{$i$} & \multicolumn{1}{|c||}{$\!\vect{\bin{i}_2\bin{i}_1}\!$} &
    \multicolumn{1}{c|}{$t_i$} \\
    \hline
    \hline
    $0$ & $\vect{\bin{0\,0}}$ & $0$ \\
    $1$ & $\vect{\bin{0\,1}}$ & $1$ \\
    $2$ & $\vect{\bin{1\,0}}$ & $1$ \\
    $3$ & $\vect{\bin{1\,1}}$ & $1$ \\
    \hline
\end{tabular}
\qquad
\begin{tabular}{|c|c|r|c||c|}
    \multicolumn{5}{c}{{Gene network} $\bs{g}_\text{dom}$} \\
    \hline
    \multicolumn{1}{|c}{$j$} & \multicolumn{1}{|c}{$\!\vect{\bin{j}_2\bin{j}_1}\!$} & \multicolumn{1}{|c}{$\mathcal{S}_j$} & \multicolumn{1}{|c||}{$k$} & \multicolumn{1}{c|}{$g_j$}\\
    \hline
    \hline
    $0$ & $\vect{\bin{0\,0}}$ & $\varnothing$ & $0$ & $3/4$ \\
    $1$ & $\vect{\bin{0\,1}}$ & $\set{1}$ & $1$ & $\!-1/4$ \\
    $2$ & $\vect{\bin{1\,0}}$ & $\set{2}$ & $1$ & $\!-1/4$ \\
    $3$ & $\vect{\bin{1\,1}}$ & $\set{1,2}$ & $2$ & $\!-1/4$ \\
    \hline
\end{tabular}
\caption{General representation of a dominant trait $\bs{t}_\text{dom}$ (\textbf{left}), and its associated gene network $\bs{g}_\text{dom}$ (\textbf{right}).} \label{Tab:2-loci_Dom_trait_gene_network}
\end{table}
\begin{figure}[!b]
\centering
\includegraphics[scale=0.55, trim={0mm 22mm 0mm 15mm},clip] % {left, bottom, right, top}
{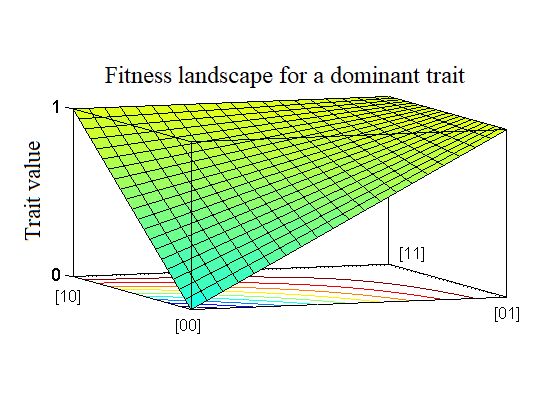}
\caption{Landscape for trait $\bs{t}_\text{dom}$ as a function of genotype $\vect{\bin{i}_2\bin{i}_1}$. The relative total roughness $\RELinflTOT{\bs{t}_\text{dom}} = 4/3$ is close to the minimal value of~$1$ in~(\ref{Eqn:REL_TOTAL_roughness_LIMITS}), so we conclude the trait landscape is \emph{fairly smooth}, which agrees with its appearance.} \label{Fig:Dom_trait}
\end{figure}

Let us calculate the local influence of the trait's roughness for the two loci from the gene network coef\mbox{}ficients. From~(\ref{Eqn:LOCAL_roughness_Fourier}) we have the influence from loci~$l=1,2$ as
\begin{eqnarray*}
\inflLOC{1}{\bs{t}_\text{dom}} &=& {\underbrace{\gr{-1/4}}_{\set{1}}}^2 + {\underbrace{\gr{-1/4}}_{\set{1,2}}}^2 \;\;=\;\; 1/8\\[4pt]
\inflLOC{2}{\bs{t}_\text{dom}} &=& {\underbrace{\gr{-1/4}}_{\set{2}}}^2 + {\underbrace{\gr{-1/4}}_{\set{1,2}}}^2 \;\;=\;\; 1/8
\end{eqnarray*}
where the underbrace for each term indicates the support set for a given cluster.
The level influences~(\ref{Eqn:LEVEL_roughness_Fourier}) for $k=0,1,2$ are
\begin{eqnarray*}
\inflLEV{0}{\bs{t}_\text{dom}} &=& 0\cdot{\underbrace{\gr{3/4}}_{\varnothing}}^2 \;\;=\;\; 0 \\[4pt]
\inflLEV{1}{\bs{t}_\text{dom}} &=& 1\cdot{\underbrace{\gr{-1/4}}_{\set{1}}}^2 \;+\; 1\cdot{\underbrace{\gr{-1/4}}_{\set{2}}}^2 \;\;=\;\; 1/8 \\[4pt]
\inflLEV{2}{\bs{t}_\text{dom}} &=& 2\cdot{\underbrace{\gr{-1/4}}_{\set{1,2}}}^2 \;\;=\;\; 1/8.
\end{eqnarray*}
Therefore, from~(\ref{Eqn:TOTAL_roughness_Fourier}), the total roughness for the dominant trait is
$\inflTOT{\bs{t}_\text{dom}} = 1/4$.

\begin{table}[!t]
\centering
\begin{tabular}{|c||c|}
    \multicolumn{2}{c}{Relative local influence} \\
    \hline
    \multicolumn{1}{|c||}{\quad$l$\;\;\;\quad} & \multicolumn{1}{c|}{$\RELinflLOC{l}{\bs{t}_\text{dom}}$} \\
    \hline
    \hline
    $1$ & $2/3$ \\
    $2$ & $2/3$ \\
    \hline
\end{tabular}
\qquad
\begin{tabular}{|c||c|}
    \multicolumn{2}{c}{Relative level influence} \\
    \hline
    \multicolumn{1}{|c||}{\quad$k$\;\;\;\quad} & \multicolumn{1}{c|}{$\RELinflLEV{k}{\bs{t}_\text{dom}}$} \\
    \hline
    \hline
    $0$ & $0$ \\
    $1$ & $2/3$ \\
    $2$ & $2/3$ \\
    \hline
\end{tabular}
\caption{The relative local influence (\textbf{left}) and relative level influence (\textbf{right}) on roughness for trait~$\bs{t}_\text{dom}$ from~(\ref{Eqn:LOCAL_roughness_Fourier}), (\ref{Eqn:LEVEL_roughness_Fourier}), (\ref{Eqn:var_Fourier}), (\ref{Eqn:REL_LOCAL_LEVEL_roughness}). The relative total roughness~(\ref{Eqn:REL_TOTAL_roughness}) is $\RELinflTOT{\bs{t}_\text{dom}} = 4/3$.} \label{Tab:2-loci_Dom_LOCAL_LEVEL_INFL}
\end{table}
The variance~(\ref{Eqn:var_Fourier}) of the trait is $\var{\bs{t}_\text{dom}} = 3/16$, and Table~\ref{Tab:2-loci_Dom_LOCAL_LEVEL_INFL}  lists the relative local and level influences on roughness~(\ref{Eqn:REL_LOCAL_LEVEL_roughness}). Examining these, we see that both loci~$l=1,2$ have equal local influence on the roughness, as do both levels~$\level{1},\level{2}$. The relative total roughness (\ref{Eqn:REL_TOTAL_roughness}) for the dominant trait is $\RELinflTOT{\bs{t}_\text{dom}} = 4/3$. As this value is closer to $1$ than $2$ (i.e., the lower, rather than the upper bound in (\ref{Eqn:REL_TOTAL_roughness_LIMITS})), we af\mbox{}firm our visual intuition that the fitness landscape of $\bs{t}_\text{dom}$ in Figure~\ref{Fig:Dom_trait} is \emph{fairly smooth}.
\end{example}

\begin{example} \label{Ex:2-loci_Inf_trait}
Our next example is a trait $\bs{t}_\text{int}$ also presented as a Boolean-valued function, where there is an ``\emph{interplay}'' between the two distinct loci.  Suppose, for instance, that the `$\bin{1}$' allele at either locus increases the concentration of a particular metabolite, where the optimum is some intermediate concentration. Then the states $\vect{\bin{10}}$ and $\vect{\bin{01}}$ would have greater fitness, while the states $\vect{\bin{00}}$ and $\vect{\bin{11}}$ are less desirable, i.e., trait values of $1$ and $0$, respectively, shown in the left side of Table~\ref{Tab:2-loci_Inf_trait_gene_network}. An idealized fitness landscape for this general sort of situation, as shown in Figure~\ref{Fig:Inf_trait}, is visibly more rugged than the landscape of~$\bs{t}_\text{dom}$. Let us see if our tools of local and level influence bear this out.

\begin{table}[!b]
\centering
\begin{tabular}{|c|c||c|}
    \multicolumn{3}{c}{{Trait} $\bs{t}_\text{int}$} \\
    \hline
    \multicolumn{1}{|c}{$i$} & \multicolumn{1}{|c||}{$\!\vect{\bin{i}_2\bin{i}_1}\!$} &
    \multicolumn{1}{c|}{$t_i$} \\
    \hline
    \hline
    $0$ & $\vect{\bin{0\,0}}$ & $0$ \\
    $1$ & $\vect{\bin{0\,1}}$ & $1$ \\
    $2$ & $\vect{\bin{1\,0}}$ & $1$ \\
    $3$ & $\vect{\bin{1\,1}}$ & $0$ \\
    \hline
\end{tabular}
\qquad
\begin{tabular}{|c|c|r|c||c|}
    \multicolumn{5}{c}{{Gene network} $\bs{g}_\text{int}$} \\
    \hline
    \multicolumn{1}{|c}{$j$} & \multicolumn{1}{|c}{$\!\vect{\bin{j}_2\bin{j}_1}\!$} & \multicolumn{1}{|c}{$\mathcal{S}_j$} & \multicolumn{1}{|c||}{$k$} & \multicolumn{1}{c|}{$g_j$}\\
    \hline
    \hline
    $0$ & $\vect{\bin{0\,0}}$ & $\varnothing$ & $0$ & $1/2$ \\
    $1$ & $\vect{\bin{0\,1}}$ & $\set{1}$ & $1$ & $0$ \\
    $2$ & $\vect{\bin{1\,0}}$ & $\set{2}$ & $1$ & $0$ \\
    $3$ & $\vect{\bin{1\,1}}$ & $\set{1,2}$ & $2$ & $\!-1/2$ \\
    \hline
\end{tabular}
\caption{General representation of trait $\bs{t}_\text{int}$ with ``interplay'' (\textbf{left}), and its associated gene network $\bs{g}_\text{int}$ (\textbf{right}).} \label{Tab:2-loci_Inf_trait_gene_network}
\end{table}
\begin{figure}[!b]
\centering
\includegraphics[scale=0.55, trim={0mm 22mm 0mm 10mm},clip] % {left, bottom, right, top}
{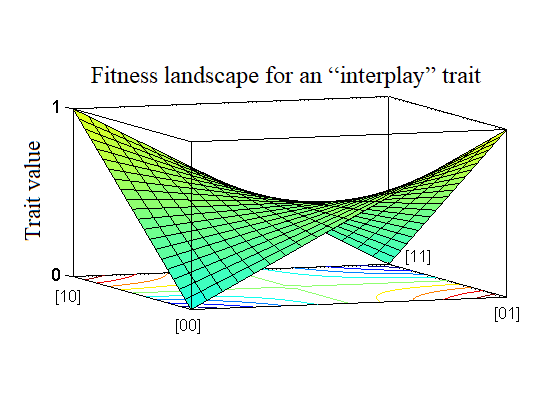}
\caption{Landscape for trait $\bs{t}_\text{int}$ as a function of genotype $\vect{\bin{i}_2\bin{i}_1}$, appears more rugged than the landscape in Fig.~\ref{Fig:Dom_trait}. The relative total roughness $\RELinflTOT{\bs{t}_\text{int}} = 2$ equals the theoretical maximum in~(\ref{Eqn:REL_TOTAL_roughness_LIMITS}), so we claim that the trait landscape is \emph{maximally rugged}.} \label{Fig:Inf_trait}
\end{figure}

The Fourier transform $\bs{g}_\text{int} = \bs{\Syl}_2\bs{t}_\text{int}/2^2$ in the right side of Table~\ref{Tab:2-loci_Inf_trait_gene_network} reveals that the level~$\level{1}$ coef\mbox{}ficients are both zero, with all of the trait's variance concentrated in the sole level~$\level{2}$ coef\mbox{}ficient.
Now the local roughnesses for loci~$l=1,2$ are
\begin{eqnarray*}
\inflLOC{1}{\bs{t}_\text{int}} &=& {\underbrace{\;\gr{0}^2}_{\set{1}}} + {\underbrace{\gr{-1/2}}_{\set{1,2}}}^2 \;\;=\;\; 1/4\\[4pt]
\inflLOC{2}{\bs{t}_\text{int}} &=& {\underbrace{\;\gr{0}^2}_{\set{2}}} + {\underbrace{\gr{-1/2}}_{\set{1,2}}}^2 \;\;=\;\; 1/4
\end{eqnarray*}
and the level influences for $k=0,1,2$ are
\begin{eqnarray*}
\inflLEV{0}{\bs{t}_\text{int}} &=& 0\cdot{\underbrace{\gr{1/2}}_{\varnothing}}^2 \;\;=\;\; 0 \\[4pt]
\inflLEV{1}{\bs{t}_\text{int}} &=& 1\cdot{\underbrace{\;\gr{0}^2}_{\set{1}}} \;+\; 1\cdot{\underbrace{\;\gr{0}^2}_{\set{2}}} \;\;=\;\; 0 \\[4pt]
\inflLEV{2}{\bs{t}_\text{int}} &=& 2\cdot{\underbrace{\gr{-1/2}}_{\set{1,2}}}^2 \;\;=\;\; 1/2
\end{eqnarray*}
resulting in a total roughness of $\inflTOT{\bs{t}_\text{int}} = 1/2$.

The variance~(\ref{Eqn:var_Fourier}) of this trait is $\var{\bs{t}_\text{int}} = 1/4$, and Table~\ref{Tab:2-loci_Inf_LOCAL_LEVEL_INFL} lists the relative local and level influences on roughness from~(\ref{Eqn:REL_LOCAL_LEVEL_roughness}). Just like Example~\ref{Ex:2-loci_Dom_trait}, both loci~$l=1,2$ have equal local influence, yet now only level~$\level{2}$ has influence on the trait's roughness (level~$\level{1}$ has zero influence). As such, the relative total roughness $\RELinflTOT{\bs{t}_\text{int}} = 2$ achieves the upper bound in~(\ref{Eqn:REL_TOTAL_roughness_LIMITS}). Hence, we can claim that the fitness landscape of~$\bs{t}_\text{int}$ in Figure~\ref{Fig:Inf_trait} is \emph{maximally rugged}. Even though this example is trivially small, we see how high-level concentration coincides with ruggedness.
\begin{table}[!h]
\centering
\begin{tabular}{|c||c|}
    \multicolumn{2}{c}{Relative local influence} \\
    \hline
    \multicolumn{1}{|c||}{\quad$l$\;\;\;\quad} & \multicolumn{1}{c|}{$\RELinflLOC{l}{\bs{t}_\text{int}}$} \\
    \hline
    \hline
    $1$ & $1$ \\
    $2$ & $1$ \\
    \hline
\end{tabular}
\qquad
\begin{tabular}{|c||c|}
    \multicolumn{2}{c}{Relative level influence} \\
    \hline
    \multicolumn{1}{|c||}{\quad$k$\;\;\;\quad} & \multicolumn{1}{c|}{$\RELinflLEV{k}{\bs{t}_\text{int}}$} \\
    \hline
    \hline
    $0$ & $0$ \\
    $1$ & $0$ \\
    $2$ & $2$ \\
    \hline
\end{tabular}
\caption{The relative local influence (\textbf{left}) and relative level influence (\textbf{right}) on roughness for trait~$\bs{t}_\text{int}$ from~(\ref{Eqn:LOCAL_roughness_Fourier}), (\ref{Eqn:LEVEL_roughness_Fourier}), (\ref{Eqn:var_Fourier}), (\ref{Eqn:REL_LOCAL_LEVEL_roughness}). The relative total roughness~(\ref{Eqn:REL_TOTAL_roughness}) is $\RELinflTOT{\bs{t}_\text{int}} = 2$.}
\label{Tab:2-loci_Inf_LOCAL_LEVEL_INFL}
\end{table}

An anonymous reviewer of this paper pointed out an interesting consequence of this situation for evolutionary pathways. The simplest evolution involves walks in the hypercube where fitness increases in a monotone fashion. The trait~$\bs{t}_\text{int}$ consists of one level~$\level{2}$ Fourier coef\mbox{}ficient, and it separates the high fitness states $\vect{\bin{01}}$ and $\vect{\bin{10}}$ into two branches, separated by a valley through $\vect{\bin{00}}$ and $\vect{\bin{11}}$.  In general, the Fourier coef\mbox{}ficient $g_{\mathcal{S}_j}$ contributes to the trait according to a multidimensional ``checkerboard pattern'' for loci in~$\mathcal{S}_j$, while it is indif\mbox{}ferent to loci outside~$\mathcal{S}_j$. Thus, when the set~$\mathcal{S}_j$ is large, a positive contribution is locally surrounded by negative contributions along every direction in~$\mathcal{S}_j$. This phenomenon tends to isolate local maxima when coef\mbox{}ficients are large and high level. Therefore, it is very intuitive to conclude that ruggedness will tend to sever monotone evolutionary paths.

But our present formalism takes no account of the distinction between homologous and heterologous loci. Many interesting examples of interplay between loci take place at two homologous loci, under the rubric ``heterozygote advantage''~\cite{BalancedPolymorph_Singh2013}. For instance, an allele causing G6PD deficiency may confer increased resistance to malaria, but also increased susceptibility to anemia. A toy model of this gives the familiar~$\bs{t}_\text{int}$ pattern\,---\,low fitness for the malaria-susceptible genotype $\vect{\bin{00}}$ and for the anemia-prone $\vect{\bin{11}}$, yet greater fitness for the heterozygotes. Such heterozygote advantage tends to preserve both alleles, rather than promote branching pathways.

Ruggedness may indeed obstruct monotone percolation paths, but it is defined as the average variability when one locus ``flips,'' without adjustment for homology of the loci.
\end{example}

% ==============================================================
% ==================================================================================
\section{Modular traits and their gene networks} \label{Sec:Modular_traits}
% ==================================================================================
% ==============================================================
Modularity is a very desirable design feature for traits, promoting resilience and evolvability. Trait modularity has profound consequences for the gene network. Modularity largely confines gene interactions to those within a module, i.e., \emph{locally}~\cite{EvolvabilityRobustness_2017}. This motivates our second governing hypothesis: due to modularity, many gene networks are, or can be approximated as sparse, with the vast majority of large coef\mbox{}ficients concentrated into the lower levels.

Consider the viability of an organism which is tested by a series of barriers to survival and reproduction. Each barrier is associated with a certain probability of successful passage. These probabilities are, to good approximation, independent and overall survival requires success with every barrier. This constitutes a sequence of $m$ filters, $f_1,f_2,\ldots,f_m$. Denote $P_i$ as the probability of surviving filter $f_i$. Then the overall survival rate is the product
\begin{equation} \label{Eqn:GauntletProcess}
P \;=\; P_1 \times P_2 \times \cdots \times P_m.
\end{equation}

These barriers represent individual modules which combine multiplicatively in the formula above. If we wish to formulate epistasis coef\mbox{}ficients as deviations from additivity, then the multiplication of probabilities leads to undesired interaction terms. The desired linear measure can be achieved by defining ``viability'' as the logarithm of the probability of surviving a suite of challenges. Hence,~(\ref{Eqn:GauntletProcess}) becomes
$$\log{P} \;=\; \log{P_1} + \log{P_2} + \cdots + \log{P_m}.$$
For instance, it has been noted in prior studies of antibiotic resistance in bacteria~\cite{EmpFitnessLandscapesPredEvol_deVisser_Krug2014} that the logarithm of survival is the appropriate version of a fitness trait, and not raw survival percentages.

A similar analysis applies to traits that are the result of multi-stage processes. The analysis of myopia genes may follow this paradigm. Here, we are concerned with survival of a focused visual image, which must endure the successive defocusing ef\mbox{}fects of the cornea, of the lens, and then the blur due to excessive axial length of the eye. If the genes causing steep cornea, dense lens, and elongated globe are in distinct modules, then their ef\mbox{}fects should be roughly additive, when quantified by diopter, the additive measure of focus. Recent work~\cite{GeneExpressionResponseOpticalDefocus_Tkatchenko2018} is beginning to identify functional modules in the genetics of myopia.

\begin{figure}[!tb]
\centering
\setlength{\unitlength}{0.75cm}

\begin{picture}(5,5)  %\thicklines
    % Locus 0
    \put(2.5,0.5){\vector(-1,1){1.995}}
    \put(1.5,1.5){\circle*{0.25}}
    \put(0.0,1.25){{\small Gene $3$}}

    % Locus 1
    \put(0.5,2.5){\vector(1,1){1.995}}
    \put(1.5,3.5){\circle*{0.25}}
    \put(0.0,3.6){{\small Gene $4$}}

    % Locus 2
    \put(2.5,0.5){\vector(1,1){1.995}}
    \put(3.5,1.5){\circle*{0.25}}
    \put(3.8,1.25){\small Gene $1$}

    % Locus 3
    \put(4.5,2.5){\vector(-1,1){1.995}}
    \put(3.5,3.5){\circle*{0.25}}
    \put(3.8,3.6){\small Gene $2$}

    % Top and bottom labels
    \put(2.05,0.05){\small \textbf{Start}}
    \put(2.0,4.75){\small \textbf{Goal}}
\end{picture}
\caption{A simple network composed of $n=4$ gene loci and two modules or branches for some generic subtask or trait. The left and right branches can be combined, e.g., as a logical \bin{AND} or \bin{OR} gate.}
\label{Fig:4-loci_Network}
\end{figure}
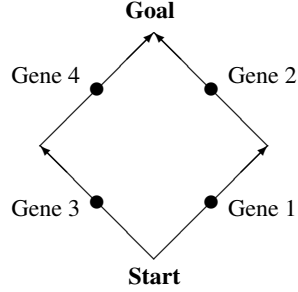

Gene loci often code for enzymes that establish a metabolic network with multiple functions. Further, these networks often resemble a logic network, although there is no exact correspondence. Under suitable restrictions of depth and size, the Fourier components of such logic networks have power spectra concentrated at low levels~\cite{ConstantDepthCircuits_Linial1993}. Metabolic networks achieve a desired state via a wide variety of genetically controlled transitions, where genes correspond to edges that permit a transition from one state to the next both in series and in parallel (see Figure~\ref{Fig:4-loci_Network}). Many processes, such as catalysis of a chemical reaction, transport across a membrane, activation of a receptor, etc., can be captured by this general structure.

% --------------------------------------------------------------
% ----------------------------------------------------------------------------------
\subsection{Very small $4$-loci modular example} \label{Sec:4-loci_Modular_example}
\begin{example} \label{Ex:4-loci_ModularTrait}
Let us use the circuit in Figure~\ref{Fig:4-loci_Network} to see how the strong interactions of the gene network are distributed due to the network's modular topology. Suppose the parallel branches are combined so as to emulate a logical \bin{OR} gate, and that the genes within each branch interact according to an \bin{AND} gate. With $n=4$ genes, there are $2^4=16$ possible genotypes. An organism, due to its genetic code, either \bin{DOES} or \bin{DOES NOT} ``have the trait,'' which we represent with a `$\bin{1}$' or `$\bin{0}$', respectively. Hence the trait value~$t_i$ for the organism with genotype $\vect{\bin{i}_4\bin{i}_3\!\bin{|}\!\bin{i}_2\bin{i}_1}$ (where the vertical bar~`$\bin{|}$' is just a visual reminder of the two branches) is
\begin{equation} \label{Eqn:SmallEx_n=4}
t_i \;=\;\Gr{\bin{i}_3 \;\bin{AND}\; \bin{i}_4} \,\;\bin{OR}\;\, \Gr{\bin{i}_1 \;\bin{AND}\; \bin{i}_2}.
\end{equation}
Table~\ref{Tab:Modular_trait_n=4} contains the full truth table for this modular trait, $\bs{t}_\text{mod}$.
\begin{table}
\centering
\begin{tabular}{|r|r||c|}
    \multicolumn{3}{c}{{Trait} $\bs{t}_\text{mod}$} \\[0pt]
    \hline
    \multicolumn{1}{|c}{$i$} & \multicolumn{1}{|c||}{$\!\vect{\bin{i}_4\bin{i}_3\!\bin{|}\!\bin{i}_2\bin{i}_1}\!$} &
    \multicolumn{1}{c|}{$t_i$} \\
    \hline
    \hline
    $ 0$ & $\vect{\bin{0 0|0 0}}$ & $0$ \\[0.125pt]
    $ 1$ & $\vect{\bin{0 0|0 1}}$ & $0$ \\[0.125pt]
    $ 2$ & $\vect{\bin{0 0|1 0}}$ & $0$ \\[0.125pt]
    $ 3$ & $\vect{\bin{0 0|1 1}}$ & ${1}$ \\[0.125pt]
    $ 4$ & $\vect{\bin{0 1|0 0}}$ & $0$ \\[0.125pt]
    $ 5$ & $\vect{\bin{0 1|0 1}}$ & $0$ \\[0.125pt]
    $ 6$ & $\vect{\bin{0 1|1 0}}$ & $0$ \\[0.125pt]
    $ 7$ & $\vect{\bin{0 1|1 1}}$ & ${1}$ \\[0.125pt]
    $ 8$ & $\vect{\bin{1 0|0 0}}$ & $0$ \\[0.125pt]
    $ 9$ & $\vect{\bin{1 0|0 1}}$ & $0$ \\[0.125pt]
    $10$ & $\vect{\bin{1 0|1 0}}$ & $0$ \\[0.125pt]
    $11$ & $\vect{\bin{1 0|1 1}}$ & ${1}$ \\[0.125pt]
    $12$ & $\vect{\bin{1 1|0 0}}$ & ${1}$ \\[0.125pt]
    $13$ & $\vect{\bin{1 1|0 1}}$ & ${1}$ \\[0.125pt]
    $14$ & $\vect{\bin{1 1|1 0}}$ & ${1}$ \\[0.125pt]
    $15$ & $\vect{\bin{1 1|1 1}}$ & ${1}$ \\
    \hline
\end{tabular}
\qquad
\begin{tabular}{|r|r|r|c||r|}
    \multicolumn{5}{c}{{Gene network} $\bs{g}_\text{mod}$} \\
    \hline
    \multicolumn{1}{|c}{$j$} & \multicolumn{1}{|c}{$\!\vect{\bin{j}_4\bin{j}_3\!\bin{|}\!\bin{j}_2\bin{j}_1}\!$} & \multicolumn{1}{|c}{$\mathcal{S}_j$} & \multicolumn{1}{|c||}{$k$} & \multicolumn{1}{c|}{$g_j$}\\
    \hline
    \hline
    $ 0$ & $\vect{\bin{0 0|0 0}}$ & $\varnothing$ & ${0}$ & $\bs{7/16}$ \\
    \hline
    $ 1$ & $\vect{\bin{0 0|0 1}}$ & $\set{1}$ & ${1}$ & $\bs{-3/16}$ \\
    $ 2$ & $\vect{\bin{0 0|1 0}}$ & $\set{2}$ & ${1}$ & $\bs{-3/16}$ \\
    $ 4$ & $\vect{\bin{0 1|0 0}}$ & $\set{3}$ & ${1}$ & $\bs{-3/16}$ \\
    $ 8$ & $\vect{\bin{1 0|0 0}}$ & $\set{4}$ & ${1}$ & $\bs{-3/16}$ \\
    \hline
    $ 3$ & $\vect{\bin{0 0|1 1}}$ & $\set{1,2}$ & ${2}$ & $ \bs{3/16}$ \\
    $ 5$ & $\vect{\bin{0 1|0 1}}$ & $\set{1,3}$ & $2$ & $-1/16$ \\
    $ 6$ & $\vect{\bin{0 1|1 0}}$ & $\set{2,3}$ & $2$ & $-1/16$ \\
    $ 9$ & $\vect{\bin{1 0|0 1}}$ & $\set{1,4}$ & $2$ & $-1/16$ \\
    $10$ & $\vect{\bin{1 0|1 0}}$ & $\set{2,4}$ & $2$ & $-1/16$ \\
    $12$ & $\vect{\bin{1 1|0 0}}$ & $\set{3,4}$ & ${2}$ & $\bs{3/16}$ \\
    \hline
    $ 7$ & $\vect{\bin{0 1|1 1}}$ & $\set{1,2,3}$ & $3$ & $1/16$ \\
    $11$ & $\vect{\bin{1 0|1 1}}$ & $\set{1,2,4}$ & $3$ & $1/16$ \\
    $13$ & $\vect{\bin{1 1|0 1}}$ & $\set{1,3,4}$ & $3$ & $1/16$ \\
    $14$ & $\vect{\bin{1 1|1 0}}$ & $\set{2,3,4}$ & $3$ & $1/16$ \\
    \hline
    $15$ & $\vect{\bin{1 1|1 1}}$ & $\set{1,2,3,4}$ & $4$ & $-1/16$ \\
    \hline
\end{tabular}
\caption{(\textbf{Left}) The modular trait $\bs{t}_\text{mod}$ resulting from the logical relationship (\ref{Eqn:SmallEx_n=4}) based on Figure~\ref{Fig:4-loci_Network}. (\textbf{Right}) The associated gene network $\bs{g}_\text{mod}$ (ordered by level index~$k$). The vertical bar~`$\bin{|}$' in $\vect{\bin{i}_4\bin{i}_3\!\bin{|}\!\bin{i}_2\bin{i}_1}$ and $\vect{\bin{j}_4\bin{j}_3\!\bin{|}\!\bin{j}_2\bin{j}_1}$ simply indicates the partition of the two modules. Notice in level~$\level{2}$ that the Fourier transform has identified the modular structure of the trait: $g_{\set{1,2}}$ and $g_{\set{3,4}}$ are $3$ times larger than $g_{\set{1,3}}$, $g_{\set{2,3}}$, $g_{\set{1,4}}$, $g_{\set{2,4}}$. Further, low-level concentration is indicated by ``cutof\mbox{}f level'' index $k_\text{cut}=2$: the interactions in levels~$\level{3}$ and~$\level{4}$ are all small.} \label{Tab:Modular_trait_n=4}
\end{table}

From~(\ref{Eqn:g=Ht}) and~(\ref{Def:Syl_Had_matrix}), the associated gene network is~$\bs{g}_\text{mod} = \bs{\Syl}_4 \bs{t}_\text{mod}/2^4$, shown in the right-hand side of Table~\ref{Tab:Modular_trait_n=4} (note, the order of~$\bs{g}_\text{mod}$ has been permuted so that its labels and coef\mbox{}ficients are grouped into their respective levels~$\set{\level{k}}_{k=0}^4$). The average of the trait, $g_0 = 7/16$, is easy to verify as there are $7$ individuals who positively have the trait. Observe that the large-magnitude (emboldened) coef\mbox{}ficients $g_j$ are in rows $j=0,1,2,4,8,3,12$, and that they occupy the lower levels $\level{k}$ for $0 \le k \le 2$. The level-ordered gene network reveals, not only low-level concentration, but the presence of a ``cutof\mbox{}f level'' index $k_\text{cut}=2$, after which we do not see any large interactions. Moreover, the only level~$\level{2}$ interactions that have significant values are the cluster pairs $\set{1,2}$ (right branch) and $\set{3,4}$ (left branch)\,---\,all other cluster pairs (i.e., $\set{1,3}$, $\set{1,4}$, $\set{2,3}$, $\set{2,4}$) are on opposite branches and have relatively small interactions\,---\,hence, \emph{the Fourier transform has been able to identify the modular structure of the trait}!

We remark that the distinction between ``large'' and ``small'' interactions of $\bs{g}$ is slight in this case (i.e., the magnitudes $7/16$ and $3/16$ versus $1/16$ in Table~\ref{Tab:Modular_trait_n=4}). However, for larger and more complicated networks, significantly greater dynamic ranges will occur, which means the gene networks will be compressible and thus well-approximated by an $\K$-sparse representation. In this sense, we can interpret the $\abs{g_j}=1/16$ coef\mbox{}ficients here as ``insignificant.'' Hence, the gene network $\bs{g}$ can be loosely characterized as ``{$7$-sparse}'' since it has $\K=7$ relatively ``large'' coef\mbox{}ficients.

It is straightforward to calculate the local and level influence on roughness in~(\ref{Eqn:LOCAL_roughness_Fourier}) and~(\ref{Eqn:LEVEL_roughness_Fourier}). Noting that the variance~(\ref{Eqn:var_Fourier}) of this modular trait is $\var{\bs{t}_\text{mod}} = 63/16^2$, the associated relative influences~$\RELinflLOC{l}{\bs{t}_\text{mod}}$ and~$\RELinflLEV{k}{\bs{t_\text{mod}}}$ in~(\ref{Eqn:REL_LOCAL_LEVEL_roughness}) are listed in Table~\ref{Tab:Modular_trait_n=4_LOCAL_LEVEL_INFL}.
As the relationship in Figure~{\ref{Fig:4-loci_Network}} and~{(\ref{Eqn:SmallEx_n=4})} are completely symmetric, we expect that the local influences from all loci $l=1,2,3,4$ to be the same. However, the influences from levels $\level{1}$ and $\level{2}$ are substantially larger than levels $\level{3}$ and $\level{4}$, which is yet another embodiment of low-level concentration. From~(\ref{Eqn:REL_TOTAL_roughness}), the relative total roughness $\RELinflTOT{\bs{t}_\text{mod}} = 96/63 \approx 1.5$ is closer to the lower limit of~$1$ than $n=4$ in~(\ref{Eqn:REL_TOTAL_roughness_LIMITS}), so this modular trait is fairly smooth.
\begin{table}[!hb]
\centering
\begin{tabular}{|c||c|}
    \multicolumn{2}{c}{Relative local influence} \\[0pt]
    \hline
    \multicolumn{1}{|c||}{\quad$l$\;\;\;\quad} & \multicolumn{1}{c|}{$\RELinflLOC{l}{\bs{t}_\text{mod}}$} \\
    \hline
    \hline
    $1$ & $24/63$ \\
    $2$ & $24/63$ \\
    $3$ & $24/63$ \\
    $4$ & $24/63$ \\
    \hline
\end{tabular}
\qquad
\begin{tabular}{|c||c|}
    \multicolumn{2}{c}{Relative level influence} \\[0pt]
    \hline
    \multicolumn{1}{|c||}{\quad$k$\;\;\;\quad} & \multicolumn{1}{c|}{$\RELinflLEV{k}{\bs{t}_\text{mod}}$} \\
    \hline
    \hline
    $0$ & $0$ \\
    $1$ & $36/63$ \\
    $2$ & $44/63$ \\
    $3$ & $12/63$ \\
    $4$ & $4/63$ \\
    \hline
\end{tabular}
\caption{The relative local influence (\textbf{left}) and relative level influence (\textbf{right}) on roughness for modular trait~$\bs{t}_\text{mod}$ from~(\ref{Eqn:LOCAL_roughness_Fourier}), (\ref{Eqn:LEVEL_roughness_Fourier}), (\ref{Eqn:var_Fourier}), (\ref{Eqn:REL_LOCAL_LEVEL_roughness}).
The symmetry in Figure~\ref{Fig:4-loci_Network} and~(\ref{Eqn:SmallEx_n=4}) results in equal local influence for loci $l=1,2,3,4$. Low-level concentration is evident since the influence from levels $\level{1}$ and $\level{2}$ is noticeably larger than from $\level{3}$ and $\level{4}$. The relative total roughness~(\ref{Eqn:REL_TOTAL_roughness}) is $\RELinflTOT{\bs{t}_\text{mod}} = 96/63 \approx 1.5$, which is fairly smooth.}
\label{Tab:Modular_trait_n=4_LOCAL_LEVEL_INFL}
\end{table}

In summary, even this very small example demonstrates the key property that we conjecture: \emph{for modular traits, the larger interactions of the gene network are confined to the lower levels}. For larger and more complicated networks, the low-level concentration ef\mbox{}fect will be much more pronounced.
\end{example}

% ----------------------------------------------------------------------------------
% --------------------------------------------------------------

% --------------------------------------------------------------
% ----------------------------------------------------------------------------------
\subsection{Simple probabilistic model}
% ----------------------------------------------------------------------------------
% --------------------------------------------------------------
\subsubsection{General ef\mbox{}fect from two modules}
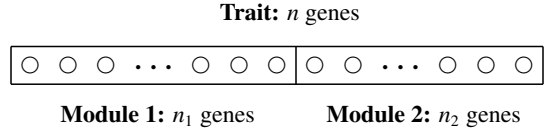
\begin{figure}[!htb]
\begin{center}
\setlength{\unitlength}{1mm}
\begin{picture}(70,20)
\linethickness{0.2mm}
\put(27.5,13.5){\small{\textbf{Trait:} $n$ genes}}
\multiput(0,5)(0,5){2}{\line(1,0){70}}
\multiput(0,5)(70,0){2}{\line(0,1){5}}
\put(37.5,5){\line(0,1){5}}
% ----
\put(6.5,0){\small{\textbf{Module 1:} $n_1$ genes}}
\multiput(2.5,7.5)(5,0){3}{\circle{2}}
\multiput(16.125,6)(2,0){3}{\Large $\cdot$}
\multiput(25,7.5)(5,0){3}{\circle{2}}
% ----
\put(41.5,0){\small{\textbf{Module 2:} $n_2$ genes}}
\multiput(40,7.5)(5,0){2}{\circle{2}}
\multiput(48.625,6)(2,0){3}{\Large $\cdot$}
\multiput(57.5,7.5)(5,0){3}{\circle{2}}
\end{picture}
\end{center}
\vspace{-10pt}
\caption{An arbitrary trait with $n$ gene loci partitioned into two modules.} \label{Fig:2-Module_Network}
\end{figure}

% --------------------------------------------------------------
Consider an arbitrary quantitative trait governed by $n$ genes, illustrated in Figure~\ref{Fig:2-Module_Network}. As previously mentioned, there are $2^n$ possible combinations in which the loci can interact. Now suppose the trait is composed of two subtasks, with $n_1$ genes in Module~$1$ dedicated to the first subtask and $n_2 = n-n_1$ genes in Module~$2$ to the second subtask. Let us assume only local interactions, where the loci of Module~$1$ do not communicate with those of Module~$2$, however this constraint can be relaxed to allow for a small amount of inter-module crosstalk.

The mere act of partitioning the genes into two modules with no cross-interactions can naturally lead to a (very) sparse gene network. To see this, let us estimate the density of significant coef\mbox{}ficients at level~$\level{k}$. The problem is isomorphic to that of calculating the probability for $k$ stones of the same color to be drawn at random from an urn with $n_1$ white stones and $n_2$ black stones, without replacement. Here, selecting $k$ stones of the same color is analogous to $k$ genes occurring in the same module. For the moment, assume $n_1$ is larger than $n_2$. Then, as $k$ increases, the white stone entries will asymptotically dominate the white-to-black ratio of successes. In fact, once $k$ exceeds~$n_2$, all successes are white, and once $k$ exceeds $n_1$, there will be no way of selecting a monocolored $k$-set. Thus there is a cutof\mbox{}f level index $k_\text{cut} = \max\gr{n_1,n_2}$.

The calculation is quite transparent if we instead return each stone to the urn after selection, i.e., if we sample with replacement. There is a probability $\alpha_1 = n_1/n$ of selecting a white stone and a corresponding $\alpha_2 = n_2/n$ of selecting a black stone, so that~$\alpha_1^k$ is the density of all-white $k$ subsets, and $\alpha_2^k$ is the density of all-black $k$ subsets. Then by analogy, $\alpha_1^k + \alpha_2^k$ bounds the density of significant coef\mbox{}ficients in level~$\level{k}$ from both Modules~$1$ and~$2$. Therefore the density of significant coef\mbox{}ficients decreases exponentially with increasing~$k$, giving a powerful impetus towards sparsity, especially in the higher levels. Notice that this argument slightly overestimates the probability of drawing $k$ stones of the same color, because it counts some cases where a stone is replaced. In turn, this can only overestimate the density of significant coef\mbox{}ficients. Yet, it can be shown with a bit more ef\mbox{}fort that the case of choosing stones without replacement yields a similar asymptotic result.

% --------------------------------------------------------------
\subsubsection{General ef\mbox{}fect from $M$ modules}
% --------------------------------------------------------------
Next, we extend the level-wise density estimate developed in the previous example from~$2$ to~${M}$ modules. For $1 \le m \le {M}$, let the $m$th module have~$n_m$ loci and probability ratio $\alpha_m = n_m/n$, with $n = \sum_m \!\:\!n_m$. Temporarily assume the loci are evenly distributed across all modules so that each $n_m = n/{M}$ and $\alpha_m = 1/{M}$. Then the density of $k$-loci clusters in any module is just $\alpha_m^k = 1/{M}^k$. Summing over all~${M}$ modules yields
\begin{equation} \label{Eqn:Prob_k-cluster_uniform}
\qquad\qquad p_k \,=\, 1/{M}^{k-1}, \qquad 1 \le k \le n.
\end{equation}
If we now account for a nonuniform distribution of loci per module, then the density at level~$\level{k}$ is simply $p_k = \sum_m \alpha_m^k$. However, asymptotically one of the $\alpha_m$ will dominate the summation, which leads to a more general form of the density, or probability that a $k$-loci cluster can significantly interact.\footnote{Notice that the density in question is among all $k$-loci clusters in the Fourier transform, and does not depend on the probability distributions of genomes in trait space. Later, we will give a proteomics example of a trait related to the anemone \emph{Entacmaea quadricolor}. The density of high-impact Fourier coef\mbox{}ficients does not require knowing the gene frequencies of the anemone, although the practical calculation does require knowing the trait value for a suf\mbox{}ficiently large set of genomes.} As $k$ increases we have the asymptotic formula
\begin{equation} \label{Eqn:Prob_k-cluster}
\qquad\qquad p_k \,\simeq\, b/a^{k-1}, \qquad 1 \le k \le n
\end{equation}
for real numbers $a>1$, $b>0$. As in the $2$-module case, there is a cutof\mbox{}f level index~$k_\text{cut} = \max\gr{\set{n_m}}$, for which there will be no way of selecting a monocolored $k$-set for $k > k_\text{cut}$.
This form can also represent the merging or melding of dif\mbox{}ferent networks. In analogy with the denominator of (\ref{Eqn:Prob_k-cluster_uniform}), we observe the parameter $a$ informally represents the ``ef\mbox{}fective number of modules.''

The number of possible $k$-loci clusters that can be drawn from $n$ loci is $\abs{\level{k}} = \binom{n}{k}$ (see~(\ref{Def:Level_Lk})). Thus for each level, multiplying $p_k$ by $\abs{\level{k}}$ yields $\K_k$, \emph{the expected number of $k$-loci clusters} that can have significant interaction energy:
\begin{equation} \label{Eqn:Sparsity_Lk}
\qquad\qquad \K_k \,=\, \Big\lfloor {p_k \cdot \abs{\level{k}}} \Big\rfloor, \qquad 1 \le k \le n
\end{equation}
where $\lfloor x \rfloor$ denotes the integer part of $x$. This can also be interpreted as \emph{the expected sparsity of level~$\level{k}$}. Note, it is always the case that $\K_0 = 1$ because there is only one element in level~$\level{0}$.

Although the asymptotic expression for $p_k$ in~(\ref{Eqn:Prob_k-cluster}) is simplistic and not likely to be exact in any real biological system, it captures the essence of the problem at hand\,---\,that modularity strongly favors interactions between fewer genes rather than many, leading to a natural concentration of significant coef\mbox{}ficients into lower levels of the gene network, resulting in sparsity. This ef\mbox{}fect is evident in~(\ref{Eqn:Sparsity_Lk}) because the polynomial growth of $\abs{\level{k}}$ cannot ``outrun'' the rate of exponential decay of~$p_k$.

% --------------------------------------------------------------
% ----------------------------------------------------------------------------------
\subsection{Sparsity as a function of the number of loci} \label{Sec:Sparsity_fnc_n}
% ----------------------------------------------------------------------------------
% --------------------------------------------------------------
Dorogovtsev and Mendes aptly point out in \cite{dorogovtsev2013evolution}, ``One should note that a number of ef\mbox{}fects in networks cannot be explained without accounting for their finite size. In this sense, most real networks are mesoscopic objects.'' Such are the gene networks we focus on. There are traits (such as certain diseases) with very small gene determinants for which our theory is unneeded, and there may well be networks so large as to be computationally beyond reach. We only claim our theory is suitable for traits in some middle zone.

Nevertheless, instead of the \kth\ level sparsity~$\K_k$ in~(\ref{Eqn:Sparsity_Lk}), it may be of interest to directly examine the asymptotic behavior of the \emph{total number of significant network coef\mbox{}ficients} $\K$ as the number of loci $n$ increases. Again assume the trait is partitioned into~$M$ modules, with~$n_m$ loci in module~$m$ and $n = \sum_m \!\:\!n_m$, and that the significant entries are only due to local interactions within each module. As before, the cutof\mbox{}f level index is the size of the largest module: $k_\text{cut} = \max\gr{\set{n_m}}$.
We want to find upper bounds that make $\K \ll 2^n$. Such bounds may be found almost \emph{ad lib}, by postulating various ways in which $k_\text{cut}$ may depend on $n$. Clearly, smaller $k_\text{cut}$ gives more stringent bounds. The following three cases cover some scenarios we may encounter:
\begin{enumerate}
\item If $k_\text{cut} < n/c$, where $1<c\le M$, then $\K$ is bounded by $c\;\!2^{n/c}$
\item If $k_\text{cut}$ is $\mathcal{O}\gr{\ln n}$, then $\K$ is bounded by a polynomial in $n$
\item If $k_\text{cut}$ is $\mathcal{O}\gr{1}$, then $\K$ is bounded by a linear function of $n$
\end{enumerate}
These estimates are all derived in a similar way, by first estimating the contribution from the largest module, and then including remaining contributions as estimated in terms of the largest component. The most interesting is Case~$2$: here, $k_\text{cut} < c \ln n$, for some constant $c$. The largest module contributes at most $2^{c\ln n}$ to $\K$, which can be rewritten as $n^{c\ln2}$. Although the number of modules $M\le n$, grossly multiplying our polynomial estimate by $n$ still leaves it polynomial, albeit one degree higher. Thus $\K$ is bounded by a polynomial in $n$.

Hence, the sparsity ratio $\K/2^n$ decays exponentially in Case~$1$, and even faster in Cases~$2$ and~$3$. Therefore, $\K \ll 2^n$ in all of these cases, as desired. The worst-case scenario of Case~$1$ is probably not very realistic, i.e., $k_\text{cut}$ most probably cannot grow without bound as a fixed proportion of~$n$. At the other extreme, the best-case scenario of Case~3 occurs when $k_\text{cut}$ has some fixed upper bound independent of~$n$, presumably due to some biological constraint. A thorough survey of empirical data from many traits is needed to ascertain if, when, and why these (or some other) cases may occur.

% --------------------------------------------------------------
% ----------------------------------------------------------------------------------
\subsection{Summary}
% ----------------------------------------------------------------------------------
% --------------------------------------------------------------
Many quantitative biological traits may appear to be simple in nature, especially when they are measured on a linear scale, but they actually represent the result of activities in multiple modular processes. This has important implications for characteristics of the trait in the Fourier domain. While modular traits may not exactly obey strict local interactions nor the asymptotic analysis above, this ideal scenario illuminates how the mechanics of modules naturally lead to:

\begin{enumerate}
\item[(i)]  A \emph{cutof\mbox{}f level index} $k_\text{cut}$ beyond which there are no, or relatively few, significant coef\mbox{}fi-cients. It follows that $k_\text{cut}$ provides a convenient way of delineating the ``low'' and ``high'' levels

\item[(ii)] \emph{Low-level concentration.} The density of significant interactions should approximately follow~(\ref{Eqn:Prob_k-cluster})

\item[(iii)] \emph{Sparsity or compressibility.} The significant entries of the gene network are a small portion of all entries
\end{enumerate}
These three properties are characteristic of traits, whether as a result of modularity or evolvability, to which compressive sensing might be profitably applied.

% ==============================================================
% ==================================================================================
\section{Compressive Sensing} \label{Sec:Compressive_Sensing}
% ==================================================================================
% ==============================================================
Although the mathematics of the two domains, the gene network space~(\ref{Eqn:g=Ht}) and trait space~(\ref{Eqn:t=Hg}), are formally symmetric, our knowledge about the two is not. We can physically measure the trait value $t_i$ of the \ith\ organism, but the scale problem means that we only have access to a very small subset of genomes. At the same time, because of the three characteristics enumerated above, we have some statistical knowledge about the distribution of all the Fourier coef\mbox{}ficients in the gene network, even without knowing the distribution of genomes in the population. Our search for Fourier coef\mbox{}ficients can be guided by the expectation that, statistically, the significant network coef\mbox{}ficients are sparse, concentrated in low levels and cut of\mbox{}f at some level. Further, from~(\ref{Eqn:t=Hg}) each trait value measurement corresponding to a particular genotype is a weighted average of the network coef\mbox{}ficients. That is, {each $t_i$ encodes partial information of the full vector~$\bs{g}$}. This setup perfectly fits the model of compressive sensing.

Compressive sensing is a combined sampling-reconstruction framework that is appropriate whenever it is expensive, or even impossible, to acquire many observations of a signal of interest. In that case, and under the correct conditions, we can take relatively few samples and still be able to reconstruct a signal with high fidelity. In the context of the genomic analysis explored in this study, we are interested in quantitative traits af\mbox{}fected by~$n$ genes or factors. The~$2^n$ combinatoric possibilities inform that it is essentially impossible to access and measure all individual organisms of a population once $n$ becomes large. Hence, compressive sensing may be an appropriate tool to permit measuring the trait values from relatively few genomes, yet still be able to analyze and quantify certain gene-to-trait functions that have been beyond realistic observable and computational means.

At the same time, some practical issues remain. The general compressive sensing model below assumes a uniform sampling, but any realistic scenario will introduce sampling bias. There exist mathematical techniques to deal with this provided that the sampling is not too biased. For example, if we sample the trait values from a population in significant linkage disequilibrium, it may af\mbox{}fect the reliability of the recovered gene network. A very simple illustration is an $n=2$-loci trait with a population equally divided between just the three genotypes $\vect{\bin{00}}$, $\vect{\bin{01}}$, $\vect{\bin{10}}$; here the linkage disequilibrium is $-1/9$. No organism with genotype $\vect{\bin{11}}$ exists, and so even exhaustive sampling can provide us only with the trait values $t_{\vect{\bin{00}}}$, $t_{\vect{\bin{01}}}$, $t_{\vect{\bin{10}}}$.

% --------------------------------------------------------------
% ----------------------------------------------------------------------------------
\subsection{Requirements of compressive sensing}
% ----------------------------------------------------------------------------------
% --------------------------------------------------------------
Two prerequisites must be met if we are to implement a compressive sensing scheme: (i) a sparse (or compressible) representation of the data of interest, and (ii) a sensing modality that is ``incoherent'' with respect to the sparsifying basis. The first condition is satisfied based on the assumed model of the traits we are interested in: those with very low roughness for the overwhelming majority of loci. The second condition is conveniently fulfilled in our model since Fourier matrices are known to be maximally incoherent relative to the standard basis, explained further below. Ultimately, this is connected to an \emph{uncertainty principle}, which dictates that localization in one domain implies its dual is ``spread out''~\cite{RobustUncPrinc_CanRomTao2006, IntroCS_Wakin2008}.\footnote{In this discrete situation, ``localized'' is synonymous with being sparse, i.e., the energy in a vector is restricted to relatively few entries, while ``spread out'' means the opposite. A simple example illustrates this: vector $\bs{g} = \vect{0,0,1,0}$ has all of its energy localized in just a single entry, whereas its Fourier transform~(\ref{Eqn:t=Hg}), $\bs{t} = \vect{1,1,-1,-1}$, has energy spread across all of its entries.} This provides a rule of thumb central to the philosophy of the compressive sensing method\,---\,by subsampling in the spread out domain (as opposed to the sparse domain) we are essentially guaranteed to gather nontrivial measurements. As the Fourier transform is a global operator, each of these measurements yield some information about the sparse domain of interest.

% --------------------------------------------------------------
% ----------------------------------------------------------------------------------
\subsection{General compressive sensing overview} \label{Sec:Compressive_Sensing_Overview}
% ----------------------------------------------------------------------------------
% --------------------------------------------------------------
We now review some key points of compressive sensing in more detail. Interested readers can find more information in the foundational and related papers, e.g., \cite{RobustUncPrinc_CanRomTao2006, CanRom_SparsityIncoherence_CS, CS_Donoho2006, CanRomTao_Noise, NearOptSigRecovFromRandProj-UnivEncStrat_CandesTao2006, SparseReconConvexRelax_FourierGauss_Vershynin2006,
RIPlessTheoryCS_CandesPlan2011}. A less formal introduction is available in dif\mbox{}ferent survey articles, such as \cite{IntroCS_Wakin2008, CS_LectureNotes_Baraniuk2007}.

Suppose we are interested in observing a real-valued $1$-D discrete signal~$\bs{g}$ of length $N$ that is sparse or compressible.\footnote{This implies that the ``sparsifying basis'' is the identity matrix.} Rather than traditional point sampling, suppose further that we acquire general linear measurements via a \emph{sensing\slash measurement matrix} $\bs{A}$ of size $M \times N$, with $M<N$. The basic compressive sensing model is embodied by the \emph{observation\slash measurement vector}
\begin{equation} \label{Eqn:y=Ag+e}
\bs{y} \,=\, \bs{Ag} + \bs{e}
\end{equation}
where $\bs{e}$ is an unknown additive noise vector of length $M$. In general, there is no hope to recover $\bs{g}$ since this is an underdetermined systems of equations (there are fewer equations than unknowns).

There are dif\mbox{}ferent ways to assemble a sensing matrix; cf.~\cite{RIPlessTheoryCS_CandesPlan2011} and the other references above. For our purposes, assume $\bs{A}$ consists of rows chosen uniformly at random from an $N \times N$ \emph{orthogonal matrix} $\bs{U}$ (i.e., $\bs{U}\bs{U}^\top\! = N\bs{I}$) that defines
the $\bs{U}$-transform of $\bs{g}$: $\bs{t} = \bs{U}\bs{g}$. As such, the observation vector $\bs{y}$ in~(\ref{Eqn:y=Ag+e}) can be thought of as an incomplete or partial sampling of a noisy transform $\bs{t}$. Define the \emph{coherence of matrix $\bs{U}$} (relative to the identity matrix)~\cite{CanRom_SparsityIncoherence_CS} as
\begin{equation} \label{Eqn:Coherence}
\mu\gr{\bs{U}} \;=\; \max_{i,j}\abs{U_{i,j}}
\end{equation}
where $1\le \mu\gr{\bs{U}} \le \!\sqrt{N}$. Highly incoherent matrices correspond to small values of $\mu$. If $\bs{g}$ is, or well-approximated as, $\K$-sparse, then with as few as
\begin{equation} \label{Eqn:M_general}
M \,=\, \text{const} \cdot \mu^2\gr{\bs{U}} \cdot \K \cdot \log{N}
\end{equation}
measurements we can estimate it, e.g., by solving the program\footnote{Alternative and equivalent formulations of (\ref{Eqn:Basis_Pursuit}) exist. Further, sFFT methods~\cite{sFFT_Indyk2014} may also be employed.}
\begin{equation} \label{Eqn:Basis_Pursuit}
\hat{\bs{g}} \;=\; \argmin_{\tilde{\bs{g}}\in\Real^N} \GR{\norm{\tilde{\bs{g}}}_1 \,\;\text{subject to}\;\, \norm{\bs{y} - \bs{A}\tilde{\bs{g}}}_2 \le \eps}
\end{equation}
where the $\ell_p$-norm of a vector $\bs{x}$ for $p=1,2$ is $\norm{\bs{x}}_p = \gr{\sum_i \abs{x_i}^p}^{1/p}$, and $\eps$ is some assumed or known measure of the energy of the noise $\bs{e}$. In words, (\ref{Eqn:Basis_Pursuit}) finds the best candidate $\tilde{\bs{g}}$ whose image under~$\bs{A}$ coincides closely with $\bs{y}$, while also being of minimal $\ell_1$-norm. The convex $\ell_1$-norm constraint is used since it is known to promote sparsity. Performance can often be improved with prior knowledge of the expected distribution of elements of~$\bs{g}$; in that case the $\norm{\tilde{\bs{g}}}_1$ regularization term can be replaced with a \emph{weighted} $\ell_1$-norm of the form $\norm{\tilde{\bs{g}}}_{1,\bs{w}} = \sum_j w_j\, \abs{\tilde{g}_j}$, with positive weights $\bs{w} = \set{w_j}$, e.g., see~\cite{Weighted_L1_Khajehnejad2009}.

\begin{remark} \label{Rem:BeautyPowerCS}
The beauty and power of compressive sensing occurs when extremely sparse $\bs{g}$ is observed via a highly incoherent sensing modality. In this case $\K \ll N$ and small~$\mu$ means we can severely undersample $\bs{t}$ with $M\ll N$ due to~(\ref{Eqn:M_general}).
\end{remark}

% --------------------------------------------------------------
% ----------------------------------------------------------------------------------
\subsection{Implications for the Fourier transform of a trait} \label{Sec:CS_Implications_traits}
% ----------------------------------------------------------------------------------
% --------------------------------------------------------------
In our model~(\ref{Eqn:t=Hg}), the sensing modality~$\bs{U}$ is the Sylvester-Hadamard matrix~$\bs{\Syl}$, which satisfies the orthogonality condition with $N=2^n$ (see~(\ref{Eqn:Syl_Orth_Relation})). Moreover, its entries are all $\pm1$. Thus~(\ref{Eqn:Coherence}) yields $\mu\gr{\bs{\Syl}} = 1$, so matrix~$\bs{\Syl}$ is \emph{maximally incoherent}. From (\ref{Eqn:M_general}), we can therefore expect to only need to observe
\begin{equation} \label{Eqn:M_trait}
M \,=\, \text{const} \cdot \K \cdot n
\end{equation}
trait values in order to accurately recover an associated gene network. Hence, the number of necessary measurements is linear in both the sparsity~$\K$ and the number of loci~$n$, whereas the number of possible genotypes is exponential in~$n$. In theory, the constant in~(\ref{Eqn:M_trait}) is small, however it is not always easy to determine it in practice. Many publications mention successful empirical studies that simply take $M\ge4 \K$~\cite{IntroCS_Wakin2008}, however these are usually associated with nonexponentially-sized vectors. Regardless of the constant factor, for large~$n$ and relatively small~$\K$ it is not unrealistic to expect~$M$ to be a tiny fraction of~$2^n$. In the next section we apply compressive sensing to a real-world trait.

% ==============================================================
% ==================================================================================
\section{An example from the literature} \label{Sec:Entacmaea_example}
% ==================================================================================
% ==============================================================
\begin{example} \label{Ex:Entacmaea_trait}
We thank an anonymous reviewer who drew our attention to a paper from Poelwijk, \emph{et al.}~\cite{LearningPatternEpistasis_Poelwijk2019}. The trait in question is the brightness of the \emph{Entacmaea quadricolor} fluorescent protein. They consider $n=13$ substitutions of one amino acid for another in the protein, and they generate all $2^{13} = 8192$ variants of the trait $\bs{t}$. While $13$ loci is still extremely small compared to many real-world traits, this is a much more meaningful example than the previous $2$- and $4$-loci examples, as there are now exponentially more genotypes and gene clusters to evaluate. Importantly, {our analysis in Sections~\ref{Sec:Discuss_Entacmaea_t_g}--\ref{Sec:Discuss_Entacmaea_power-law_density} is completely in the Fourier domain}\,---\,either directly ``reading of\mbox{}f'' gene network coef\mbox{}ficients or combinations of their energies.

% --------------------------------------------------------------
% ----------------------------------------------------------------------------------
\subsection{Discussion of the full trait and gene network} \label{Sec:Discuss_Entacmaea_t_g}
% ----------------------------------------------------------------------------------
% --------------------------------------------------------------
The completely measured trait $\bs{t}$ is plotted as function of decimal genotype index~$i$ on the left side of Figure~\ref{Fig:Entacmaea_t_g_gLevel}. Using~(\ref{Eqn:g=Ht}) and~(\ref{Def:Syl_Had_matrix}) with $n=13$, we can compute all $8192$ Fourier coef\mbox{}ficients: $\bs{g} = \bs{\Syl}_{\!13}\bs{t}/2^{13}$, which is
plotted as function of decimal cluster index~$j$ in the upper-right of Figure~\ref{Fig:Entacmaea_t_g_gLevel}. Right away we see that the average value of the trait in coef\mbox{}ficient $g_0$ is slightly larger than $0.5$. This makes sense, as the bulk of the trait values are close to $0.3$. Next we observe that there are relatively few large coef\mbox{}ficients (the ten largest are identified with colored dots) in the gene network, along with a handful of medium-small sized values, with the rest being very small and noise-like; thus the vector~$\bs{g}$ appears to be quite compressible.
But does the gene network possess the desirable property of \emph{low-level concentration}? The lower-right plot of Figure~\ref{Fig:Entacmaea_t_g_gLevel} shows the gene network's coef\mbox{}ficients permuted so that they are ordered according to their respective levels.\footnote{Within each level the indices follow obvious ordering: in level~$\level{1}$ the first indices are $j=1,2,4,8,\ldots$, in level~$\level{2}$ the first indices are $j=3,5,6,9,\ldots$, in level~$\level{3}$ the first indices are $j=7,11,13,14,\ldots$, and so on.} From inspection, the largest coef\mbox{}ficients clearly fall within lower levels, with the ``heaviest hitters,'' including the top ten, concentrated into levels~$\level{0}$--$\level{3}$. As mentioned in the Introduction, this is analogous to a traditional signal dominated by low frequencies, rather than high.
\begin{figure}[!tb]
\centering
~{}\hspace{-27pt}\includegraphics[scale=0.35, trim={45mm 10mm 0mm 5mm},clip] % {left, bottom, right, top}
{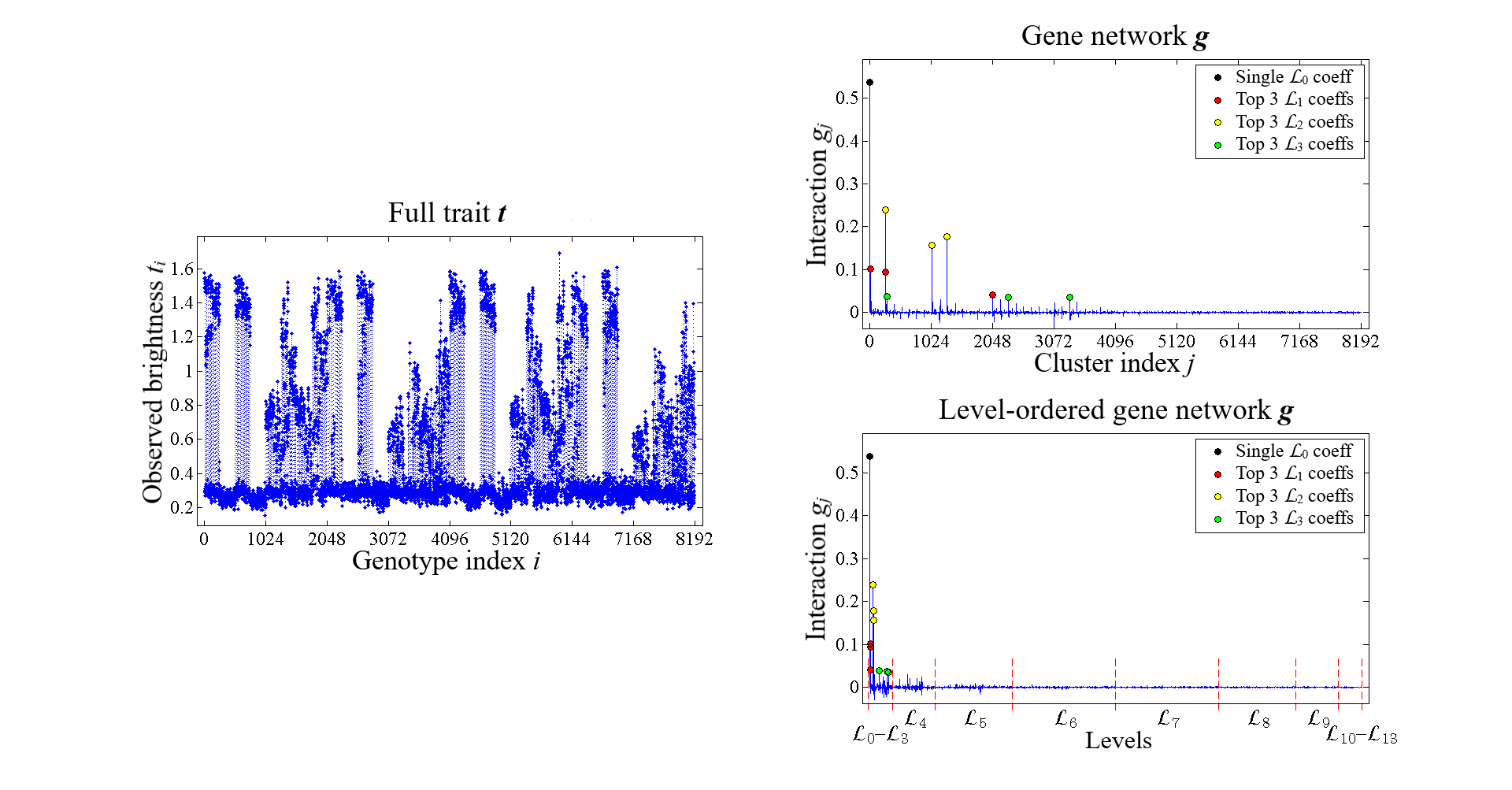}
\caption{(\textbf{Left}) Observed trait data~$\bs{t}$ of the \emph{Entacmaea quadricolor} fluorescent protein for all $N=2^{13}=8192$ genotypes (from~\cite{LearningPatternEpistasis_Poelwijk2019}). (\textbf{Top-right}) The associated gene network of interactions~$\bs{g}$ from~(\ref{Eqn:g=Ht}). The colored dots indicate the ten largest-magnitude coef\mbox{}ficients. (\textbf{Bottom-right}) Coef\mbox{}ficients of the same gene network but reordered into their respective levels. Notice how the largest interactions (including the top ten) are concentrated into the lower levels. In terms of conventional Fourier analysis, this is analogous to a low-frequency signal.} \label{Fig:Entacmaea_t_g_gLevel}
\end{figure}

Let us now examine, say, the ten largest-magnitude interactions of the gene network, listed in Table~\ref{Tab:Entacmaea_top-10_interactions}. Note, the energy of these ten coef\mbox{}ficients (i.e., sum of squares of just these $g_j$) is $0.4277$, and the total energy (i.e., sum of squares of all $g_j$) is $0.4487$, so these very few interactions already capture $95.3\%$ of the trait's expression.
For each coef\mbox{}ficient~$g_j$ (colored dot in the upper-right panel of Figure~\ref{Fig:Entacmaea_t_g_gLevel}), its index~$j$ indicates the location in the decimal-ordered vector. Converting $j$ to its equivalent $13$-bit string $\vect{\bin{j}_{13}\ldots\bin{j}_2\!\:\bin{j}_1}$ and support set~$\mathcal{S}_j$ reveals which loci are members of the \jth\ cluster being evaluated. After the trivial coef\mbox{}ficient $g_0$, we see that the three largest meaningful interactions all consist of pairs of loci, the next three are all singletons, and the next three are all triads. Notice that the cluster pairs $\set{4,9},\set{9,11},\set{4,11}$ form the edges of a triangle for the loci~$l=4,9,11$. The cluster singletons $\set{4},\set{9}$ appear right after as meaningful, yet singleton $\set{12}$ shows up next, and $\set{11}$ is not even in the top ten. Further, the cluster triad $\set{4,9,11}$ shows up as thirty-second in the list of sorted descending magnitude coef\mbox{}ficients.

\begin{table}[!tb]
\centering
\setlength\tabcolsep{3.0pt}
\begin{tabular}{|r|ccccccccccccc|r|c||c|c|}
    \multicolumn{18}{c}{{Top ten coef\mbox{}ficients of gene network}, $\bs{g}$} \\

    \hline

    \multicolumn{1}{|c|}{$j$} &
    $\,[\bin{j}_{13}\!$ & $\bin{j}_{12}$ & $\bin{j}_{11}$ & $\bin{j}_{10}$ & $\bin{j}_{9\;}$ & $\bin{j}_{8\;}$ & $\bin{j}_{7\;}$ & $\bin{j}_{6\;}$ & $\bin{j}_{5\;}$ & $\bin{j}_{4\;}$ & $\bin{j}_{3\;}$ & $\bin{j}_{2\;}$ & $\bin{j}_1]\,$
    & \multicolumn{1}{c}{$\mathcal{S}_j$} & \multicolumn{1}{|c||}{$\;k\;$} & \multicolumn{1}{c|}{$\quad\: g_j \quad\:$} & \multicolumn{1}{c|}{\,\tikz\draw[black] (0,0) circle (.35ex);\,}\\

    \hline
    \hline

    $0\,$ & $\![\,\bin{0}$ & $\bin{0}$ & $\bin{0}$ & $\bin{0}$ & $\bin{0}$ & $\bin{0}$ & $\bin{0}$ & $\bin{0}$ & $\bin{0}$ & $\bin{0}$ & $\bin{0}$ & $\bin{0}$ & $\,\bin{0}\,]$ & $\varnothing\,$ & $0$ & $0.5381$ & \tikz\draw[black,fill=black] (0,0) circle (.35ex);\\

    $264\,$ & $\![\,\bin{0}$ & $\bin{0}$ & $\bin{0}$ & $\bin{0}$ & $\bin{1}$ & $\bin{0}$ & $\bin{0}$ & $\bin{0}$ & $\bin{0}$ & $\bin{1}$ & $\bin{0}$ & $\bin{0}$ & $\,\bin{0}\,]$ & $\set{4,9}\,$ & $2$ & $\bs{0.2396} $ & \tikz\draw[black,fill=yellow] (0,0) circle (.35ex);\\

    $1280\,$ & $\![\,\bin{0}$ & $\bin{0}$ & $\bin{1}$ & $\bin{0}$ & $\bin{1}$ & $\bin{0}$ & $\bin{0}$ & $\bin{0}$ & $\bin{0}$ & $\bin{0}$ & $\bin{0}$ & $\bin{0}$ & $\,\bin{0}\,]$ & $\set{9,11}\,$ & $2$ & $\bs{0.1778}$ & \tikz\draw[black,fill=yellow] (0,0) circle (.35ex);\\

    $1032\,$ & $\![\,\bin{0}$ & $\bin{0}$ & $\bin{1}$ & $\bin{0}$ & $\bin{0}$ & $\bin{0}$ & $\bin{0}$ & $\bin{0}$ & $\bin{0}$ & $\bin{1}$ & $\bin{0}$ & $\bin{0}$ & $\,\bin{0}\,]$ & $\set{4,11}\,$ & $2$ & $\bs{0.1565}$ & \tikz\draw[black,fill=yellow] (0,0) circle (.35ex);\\

    $8\,$ & $\![\,\bin{0}$ & $\bin{0}$ & $\bin{0}$ & $\bin{0}$ & $\bin{0}$ & $\bin{0}$ & $\bin{0}$ & $\bin{0}$ & $\bin{0}$ & $\bin{1}$ & $\bin{0}$ & $\bin{0}$ & $\,\bin{0}\,]$ & $\set{4}\,$ & $1$ & $0.1019$ &
    \tikz\draw[black,fill=red] (0,0) circle (.35ex);\\

    $256\,$ & $\![\,\bin{0}$ & $\bin{0}$ & $\bin{0}$ & $\bin{0}$ & $\bin{1}$ & $\bin{0}$ & $\bin{0}$ & $\bin{0}$ & $\bin{0}$ & $\bin{0}$ & $\bin{0}$ & $\bin{0}$ & $\,\bin{0}\,]$ & $\set{9}\,$ & $1$ & $0.0934$ & \tikz\draw[black,fill=red] (0,0) circle (.35ex);\\

    $2048\,$ & $\![\,\bin{0}$ & $\bin{1}$ & $\bin{0}$ & $\bin{0}$ & $\bin{0}$ & $\bin{0}$ & $\bin{0}$ & $\bin{0}$ & $\bin{0}$ & $\bin{0}$ & $\bin{0}$ & $\bin{0}$ & $\,\bin{0}\,]$ & $\set{12}\,$ & $1$ & $0.0401$ & \tikz\draw[black,fill=red] (0,0) circle (.35ex);\\

    $280\,$ & $\![\,\bin{0}$ & $\bin{0}$ & $\bin{0}$ & $\bin{0}$ & $\bin{1}$ & $\bin{0}$ & $\bin{0}$ & $\bin{0}$ & $\bin{1}$ & $\bin{1}$ & $\bin{0}$ & $\bin{0}$ & $\,\bin{0}\,]$ & $\set{4,5,9}\,$ & $3$ & $0.0380$ & \tikz\draw[black,fill=green] (0,0) circle (.35ex);\\

    $2312\,$ & $\![\,\bin{0}$ & $\bin{1}$ & $\bin{0}$ & $\bin{0}$ & $\bin{1}$ & $\bin{0}$ & $\bin{0}$ & $\bin{0}$ & $\bin{0}$ & $\bin{1}$ & $\bin{0}$ & $\bin{0}$ & $\,\bin{0}\,]$ & $\set{4,9,12}\,$ & $3$ & $0.0360$ & \tikz\draw[black,fill=green] (0,0) circle (.35ex);\\

    $\,3328\,$ & $\![\,\bin{0}$ & $\bin{1}$ & $\bin{1}$ & $\bin{0}$ & $\bin{1}$ & $\bin{0}$ & $\bin{0}$ & $\bin{0}$ & $\bin{0}$ & $\bin{0}$ & $\bin{0}$ & $\bin{0}$ & $\,\bin{0}\,]$ & $\,\set{9,11,12}\,$ & $3$ & $0.0352$ & \tikz\draw[black,fill=green] (0,0) circle (.35ex);\\

    \hline
\end{tabular}
\caption{The top ten magnitude coef\mbox{}ficients of the gene network $\bs{g}$ and their associated colored dots in the upper-right plot of Fig.~\ref{Fig:Entacmaea_t_g_gLevel}. The index $j$ indicates their location in the decimal-ordered vector. The participating loci in each interaction are identified by either the equivalent binary string $\vect{\bin{j}_{13}\ldots\bin{j}_1}$ or support set~$\mathcal{S}_j$. The level index~$k$ is the cardinality of the support set. These ten low-level clusters contain $95.3\%$ of the energy of all interactions. The three level~$\level{2}$ interactions (bold) have the most influence of roughness seen in Fig.~\ref{Fig:Entacmaea_Local_Level_roughness_relative}.} \label{Tab:Entacmaea_top-10_interactions}
\end{table}

%\newpage{}~ %% Can remove this if causing typesetting problems in final journal version
%\newpage
There is no strict definition of ``low-level concentration,'' but a reasonable approach is to examine how the top interactions of a gene network are distributed across its levels. A sorted descending order of all~$8192$ interaction magnitudes (not shown) does not have an obvious breakpoint to indicate which are the strongest. Figure~\ref{Fig:Entacmaea_dist_low-level_interactions} shows how the largest $\K = 200$ interactions contained in $\bs{g}$ are distributed; we observe that the vast majority of significant interactions are in level~$\level{5}$ and lower.\footnote{It is worth pointing out that the histogram in Figure~\ref{Fig:Entacmaea_dist_low-level_interactions} is a bit misleading since many of the top $200$ interactions are actually quite small. To see this, refer to level-ordered gene network $\bs{g}$ in the bottom-right of Figure~\ref{Fig:Entacmaea_t_g_gLevel}, and notice that the coef\mbox{}ficients in levels~$\level{5}$--$\level{7}$ are extremely small, yet $30$--$40$ of these are significant enough to be included in the top~$200$.} As a percentage of the $8192$ total interactions in the gene network, these $200$ represent just $2.44\%$ of the entries of~$\bs{g}$. Yet, at the same time, this small collection of clusters has an energy of $0.4449$, so they capture an impressive $99.2\%$ of the energy possessed by gene network.
This demonstrates that the gene network's meaningful interactions are confined to: (i) relatively few clusters, and (ii) these clusters contain relatively few loci\,---\,this exemplifies the phenomenon of low-level concentration. For the traits we are interested in that are governed by a larger number of loci~$n$, we expect to see an even more pronounced concentration of meaningful clusters into the lower levels, resulting in small $\K$ relative $2^n$. Nonetheless, let us provisionally take the working sparsity as $\K = 200$ meaningful interactions in~$\bs{g}$.

\begin{figure}[!b]
\centering
\includegraphics[scale=0.35, trim={10mm 0mm 0mm 0mm},clip] % {left, bottom, right, top}
{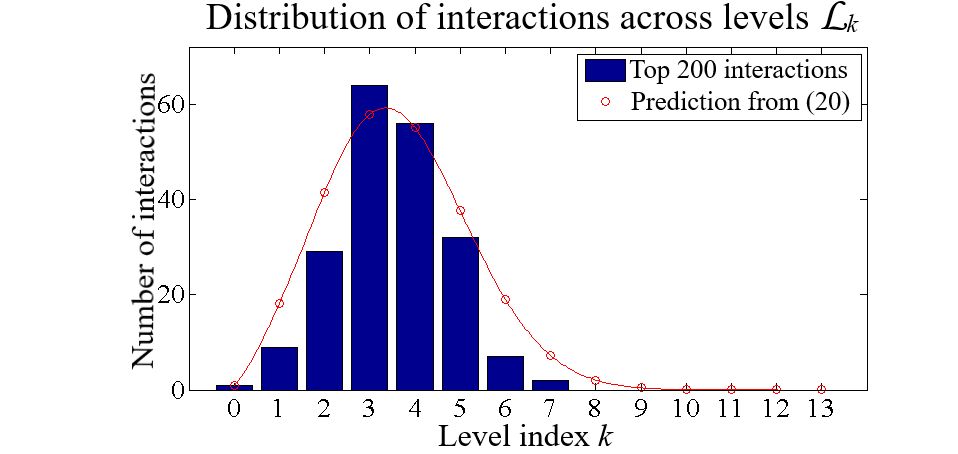}
\caption{Evidence that the strongest $\K = 200$ gene network interactions in Fig.~\ref{Fig:Entacmaea_t_g_gLevel} are concentrated into the lower levels. Even though these interactions are only $2.44\%$ of the entries of~$\bs{g}$, they contain $99.2\%$ of the gene network's energy. The red circles show a satisfactory prediction from (\ref{Eqn:Sparsity_Lk}).}
\label{Fig:Entacmaea_dist_low-level_interactions}
\end{figure}

As shown in Figure~\ref{Fig:Entacmaea_dist_low-level_interactions}, the top interaction terms have a unimodal distribution that declines rapidly past levels~$\level{3}$ and~$\level{4}$. Figure~\ref{Fig:Entacmaea_dist_low-level_interactions} also displays the prediction of the theoretical level-sparsity~$\K_k$~(\ref{Eqn:Sparsity_Lk}).
Even though derived from combinatorial considerations, it still gives a good feel for the general shape of the distribution of the largest Fourier coef\mbox{}ficients, at least for this one real-world trait. The parameters used in~(\ref{Eqn:Sparsity_Lk}) were $a=2.4$ and $b=1.4$. Informally, we can interpret $a$ here as ``$2.4$ ef\mbox{}fective modules,'' which in turn means there are, on average, $5.4$ loci per module. This is not unreasonable as the distribution of the real data (blue bars) in Figure~\ref{Fig:Entacmaea_dist_low-level_interactions} shows meaningful interactions confined to  level~$\level{7}$ and below.

It is dif\mbox{}ficult to discern a modular structure in the brightness trait data. As noted in Section~\ref{Sec:Compare:CoexpressionNetsModules}, it may be dif\mbox{}ficult to detect modularity in noisy data. Nevertheless, the present example is much too small to support more complicated predictions. Further empirical data is needed to determine whether the low-level concentration encapsulated in~(\ref{Eqn:Sparsity_Lk}), whether due to modularity or simple evolvability, is pervasive.

% --------------------------------------------------------------
% ----------------------------------------------------------------------------------
\subsection{Discussion of factors af\mbox{}fecting the distribution of roughness}
% ----------------------------------------------------------------------------------
% --------------------------------------------------------------
With $n=13$ loci, the distinction between the local and level influences on roughness becomes much more meaningful than in Examples~\ref{Ex:2-loci_Dom_trait}--\ref{Ex:4-loci_ModularTrait}. From the gene network~$\bs{g}$ in Figure~\ref{Fig:Entacmaea_t_g_gLevel} we calculate $\inflLOC{l}{\bs{t}}$ in~(\ref{Eqn:LOCAL_roughness_Fourier}) and $\inflLEV{k}{\bs{t}}$ in~(\ref{Eqn:LEVEL_roughness_Fourier}); the variance in~(\ref{Eqn:var_Fourier}) is $\sigma^2(\bs{t}) = 0.16$. From~(\ref{Eqn:LOCAL_roughness_Fourier}), (\ref{Eqn:LEVEL_roughness_Fourier}), (\ref{Eqn:var_Fourier}),
the associated relative influences~$\RELinflLOC{l}{\bs{t}}$ and~$\RELinflLEV{k}{\bs{t}}$ in~(\ref{Eqn:REL_LOCAL_LEVEL_roughness}) are seen in Figure~\ref{Fig:Entacmaea_Local_Level_roughness_relative}. In the left-hand plot, we immediately see that loci $l=4,9,11$ exert the most local influence on the trait's roughness. This is borne out in the right-hand plot with the majority of the level influence focused into level~$\level{2}$. Returning to Table~\ref{Tab:Entacmaea_top-10_interactions} it is evident that these influences are mostly due to the first three interactions (seen in bold) in cluster pairs $\set{4,9},\set{9,11},\set{4,11}$.

The relative total roughness of $\RELinflTOT{\bs{t}} = 2.1$ from~(\ref{Eqn:REL_TOTAL_roughness}) is much closer to the minimum value of~$1$ rather than the maximum of $n=13$ in~(\ref{Eqn:REL_TOTAL_roughness_LIMITS}), so we can confidently claim that this trait has a rather smooth landscape. A relatively small value of $\RELinflTOT{\bs{t}}$ is another indication of low-level concentration, since if there had been meaningful interactions in the higher levels, then they would have swamped the influences due to their $k$-fold presence. This pattern is one of a few highly influential gene loci with extensive interactions and many loci with small influence. Assuming the trait is already optimized, the few highly influential loci would be highly resistant to substitution, while the many low-influence loci are more susceptible to modification.
\begin{figure}[!bt]
\centering
\includegraphics[scale=0.325, trim={45mm 0mm 0mm 0mm},clip] % {left, bottom, right, top}
{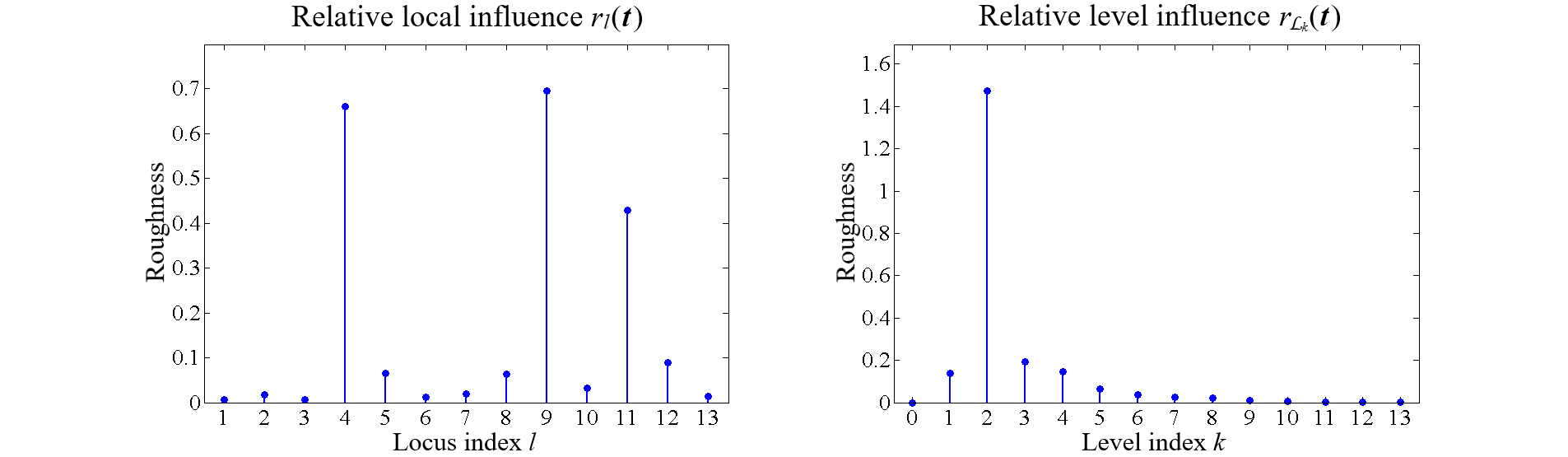}
\caption{(\textbf{Left}) The relative local influence on roughness $\RELinflLOC{l}{\bs{t}}$ shows that loci~$l=4,9,11$ exert the most influence on the trait. (\textbf{Right}) The relative level influence on roughness $\RELinflLEV{k}{\bs{t}}$ reveals that level~$\level{2}$ has a disproportionate impact. Compare with the gene interactions in Table~\ref{Tab:Entacmaea_top-10_interactions}. The relative total roughness, $\RELinflTOT{\bs{t}} = 2.1$, indicates that the overall trait landscape is very smooth.} \label{Fig:Entacmaea_Local_Level_roughness_relative}
\end{figure}

% --------------------------------------------------------------
% ----------------------------------------------------------------------------------
\subsection{Discussion of density of local influence} \label{Sec:Discuss_Entacmaea_power-law_density}
% ----------------------------------------------------------------------------------
% --------------------------------------------------------------
A scale-free network is characterized by just this pattern of many elements of low connectivity and a few of high connectivity, following a power-law distribution. Barab\'asi and Albert drew attention to processes whereby the scale-free distribution emerges in evolving networks due to a preferential attachment mechanism~\cite{EvolutionNetworks_Dorogovtsev2002}. Such networks are relatively fault tolerant.
We expect fault tolerance and we expect that gene network evolution has involved accretion of novel genetic material with subsequent modification (analogous to the Barab\'asi-Albert evolution). It is a natural question, although dif\mbox{}ficult to answer, whether gene networks are scale-free using roughness as an analogue of valency in graphs.

The goal is to show that the density of local influence on roughness roughly follows a power law. To determine this we first arrange the local influences on roughness, $\RELinflLOC{l}{\bs{t}}$ (left side of Figure~\ref{Fig:Entacmaea_Local_Level_roughness_relative}), in their ascending sorted order, seen on the left side of Figure~\ref{Fig:Entacmaea_localRoughnessDensity-Barabarsi} as blue dots, along with their best fit curve in red; this shows the local influence on roughness as a function of sorted index. However, we need to quantify \emph{loci count as a function of roughness}. As such, the blue dots and red curve in the plot on the right side of the figure show the inverses of the functions in the left plot. Thus, the red curve is the function which approximately counts the number of loci below a given roughness value. For instance, there are~$10$ loci with (influence on) roughness less~$0.3$.

The \emph{density of loci of a certain roughness} is simply the derivative with respect to roughness of the function that counts the number of loci up to that roughness. The black curve in the right plot of Figure~\ref{Fig:Entacmaea_localRoughnessDensity-Barabarsi} shows this best fit to density: empirically the equation is $0.28x^{-0.79}$, with $x$ denoting roughness. Notice how it captures the high density of the blue dots for loci with small influence on roughness, and very low density (i.e., rareness) of large influences. This density follows the desired power-law decay, which suggests a scale-free network, but requires further investigation, as little can be concluded from such scant data. In general, it is very dif\mbox{}ficult to establish a power-law structure even for large, evolving networks such as the internet.
\begin{figure}[!tb]
\centering
\includegraphics[scale=0.325, trim={50mm 0mm 0mm 0mm},clip] % {left, bottom, right, top}
{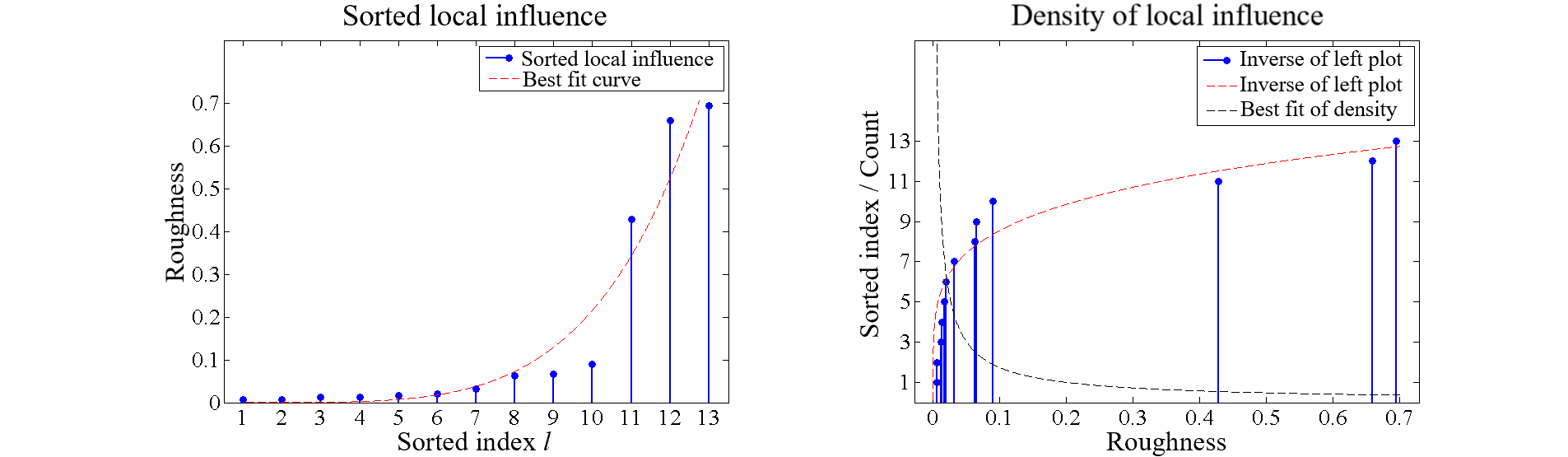}
\caption{(\textbf{Left}) The local influences on roughness from the left plot of Fig.~\ref{Fig:Entacmaea_Local_Level_roughness_relative} sorted in ascending order. (\textbf{Right}) The blue dots and red curve are the inverse functions of those seen in the left plot. The red curve on the right is an approximation to the number of loci that have (influence) less than a given roughness. The black curve is the derivative of the red curve and approximates the \emph{density of loci} with (influence of) a given roughness. This density has a power-law decay: $0.28x^{-0.79}$, where $x$ denotes roughness.} \label{Fig:Entacmaea_localRoughnessDensity-Barabarsi}
\end{figure}

% --------------------------------------------------------------
% ----------------------------------------------------------------------------------
\subsection{Using compressive sensing to subsample and reconstruct}
% ----------------------------------------------------------------------------------
% --------------------------------------------------------------
Armed with the knowledge that this proteomic trait is smooth and therefore sparse, we can apply compressive sensing techniques to subsample the trait values and reconstruct its associated gene network. Note, our results largely confirm those reported in~\cite{LearningPatternEpistasis_Poelwijk2019}. From the theory presented in Sections~\ref{Sec:Compressive_Sensing_Overview} and~\ref{Sec:CS_Implications_traits}, suppose we have access to only $M=819$ of the $N=8192$ possible genotypes, a subsampling ratio of $M/N=10\%$, and that these individuals represent a uniformly random sampling from the Boolean cube. Note, with our \emph{a priori} knowledge that $\bs{g}$ is well-approximated as $\K$-sparse with $\K=200$, we expect, since this is not such a large data set, a good reconstruction from $M=819$ samples as it is more than $4s$.
Measure each individual's trait value $t_i$ and record it as $y_i$ in the partial trait/observation vector~$\bs{y}$, shown in Figure~\ref{Fig:y_Observ}; compare the $10\%$ subsampling with the full trait~$\bs{t}$ in Figure~\ref{Fig:Entacmaea_t_g_gLevel}. From the genotypes of the observed trait values, we assemble our sensing matrix~$\bs{A}$ as the associated rows of the full Sylvester-Hadamard matrix~$\bs{\Syl}$.
\begin{figure}[!tb]
\centering
\includegraphics[scale=0.35, trim={0mm 70mm 0mm 60mm},clip] % {left, bottom, right, top}
{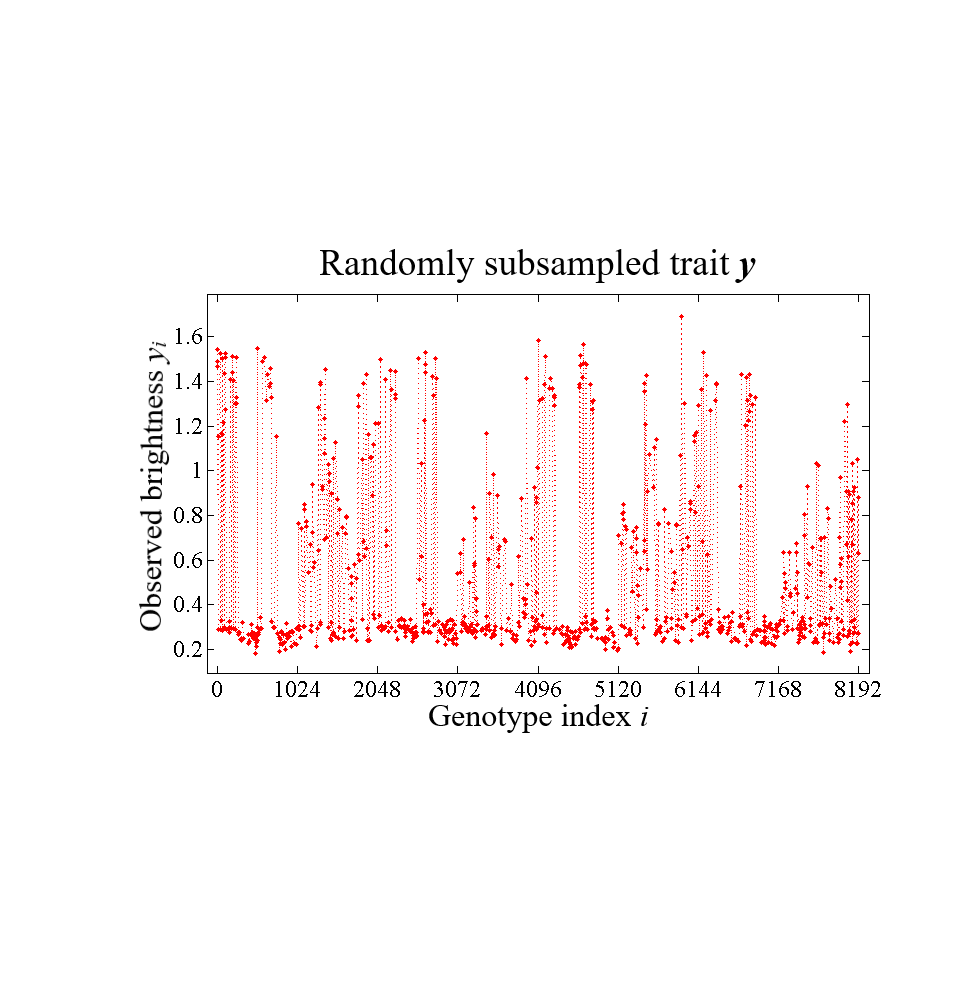}
\caption{Partial trait/observation vector $\bs{y}$ with the trait values of just $M=819$ randomly chosen individuals. Compare the $10\%$ subsampling with the full trait's $N=8192$ values in Fig.~\ref{Fig:Entacmaea_t_g_gLevel}.} \label{Fig:y_Observ}
\end{figure}

\begin{figure}[!b]
\centering
\includegraphics[scale=0.35, trim={0mm 10mm 0mm 9mm},clip] % {left, bottom, right, top}
{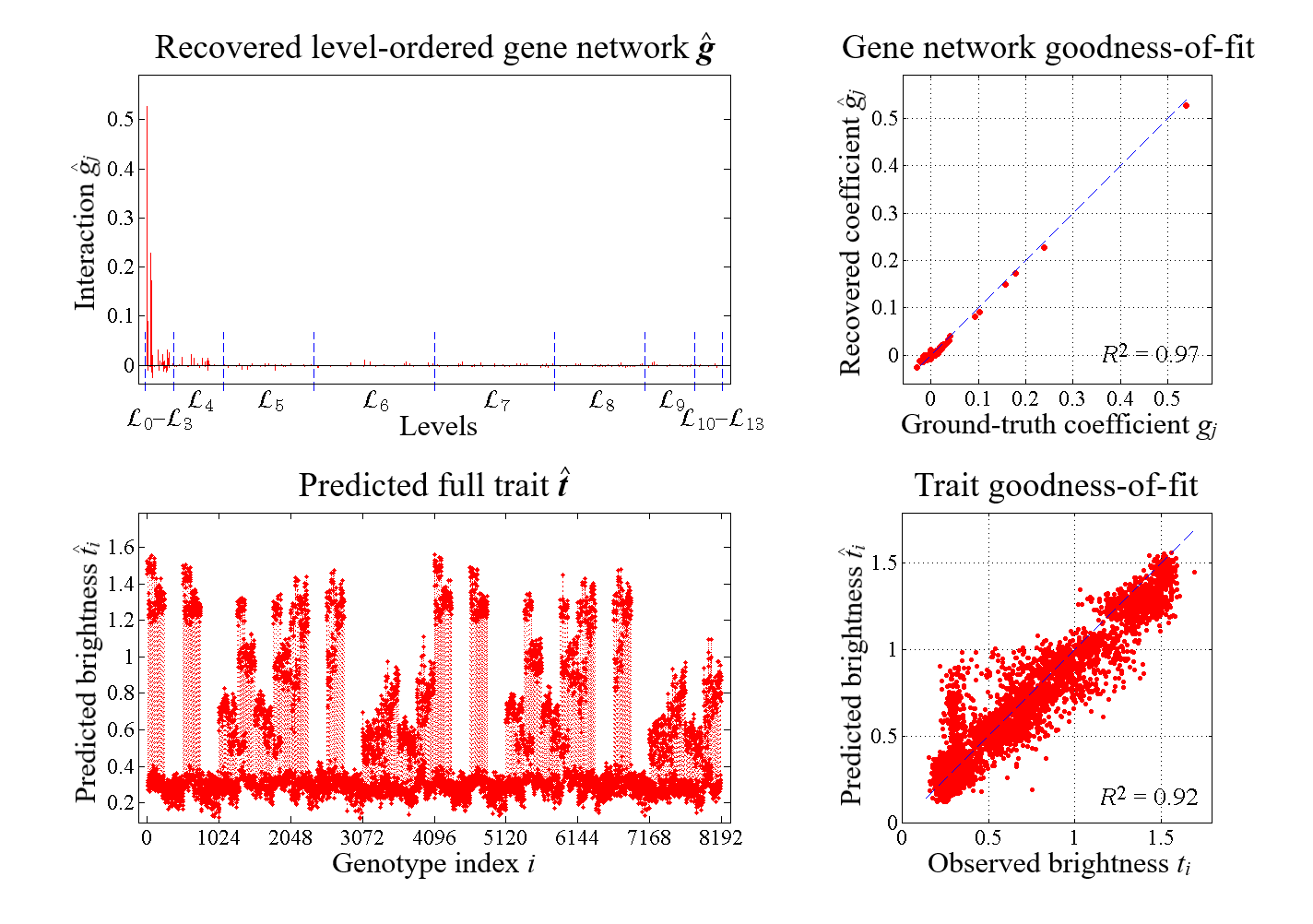}
\caption{(\textbf{Top-left}) The reconstructed gene network~$\hat{\bs{g}}$ with its top $\K=200$ interactions obtained from~(\ref{Eqn:Basis_Pursuit}) using the $M=819$ ($10\%$ subsampling) randomly observed trait values~$\bs{y}$ seen in Fig.~\ref{Fig:y_Observ}. Compare with the ground truth~$\bs{g}$ in the bottom-right of Fig.~\ref{Fig:Entacmaea_t_g_gLevel}. (\textbf{Top-right}) The goodness-of-fit for the recovered top $\K=200$ coef\mbox{}ficients~$\set{\hat{g}_j}$ relative to their ground truths~$\set{g_j}$ is excellent: $R^2 = 0.97$. (\textbf{Bottom-left}) The predicted full trait~$\hat{\bs{t}}$ for all genotypes from the inverse Fourier transform of~$\hat{\bs{g}}$. Compare with the ground truth~$\bs{t}$ in the left side of Fig.~\ref{Fig:Entacmaea_t_g_gLevel}. (\textbf{Bottom-right}) The goodness-of-fit for the predicted full trait~$\hat{\bs{t}}$ relative to the observed trait~$\bs{t}$ is very good: $R^2 = 0.92$.} \label{Fig:Entacmaea_CS_recovery}
\end{figure}

Given measurements $\bs{y}$ and matrix $\bs{A}$, we recovered an approximate gene network $\hat{\bs{g}}$ by implementing (\ref{Eqn:Basis_Pursuit}) using the ``Fast Adaptive Shrinkage/Thresholding Algorithm'' (FASTA)~\cite{FASTA:2014}. The upper-left plot of Figure~\ref{Fig:Entacmaea_CS_recovery} shows the top $\K=200$ coef\mbox{}ficients of the recovered gene network $\hat{\bs{g}}$ in red. The gene network is displayed in its permuted level order. Clearly, the large-magnitude coef\mbox{}ficients in the lower levels were faithfully recovered, as verified in the upper-right plot, which shows the goodness-of-fit relative to the ground truth; the coef\mbox{}ficient of determination $R^2 = 0.97$ is very good (see the ground-truth level-ordered gene network $\bs{g}$ in the bottom-right plot of Figure~\ref{Fig:Entacmaea_t_g_gLevel}). Notice that FASTA also recovered some very small magnitude interactions that are located in the higher levels. These can be suppressed and the lower levels favored by using a weighted $\ell_1$-norm, as mentioned in Section~\ref{Sec:Compressive_Sensing_Overview}.

Note, the full trait is at our disposal in this example, so the ground-truth gene network~$\bs{g}$ was known in Sections~\ref{Sec:Discuss_Entacmaea_t_g}--\ref{Sec:Discuss_Entacmaea_power-law_density}. We certainly do not expect this in most situations. However, once a suf\mbox{}ficiently reliable reconstructed gene network~$\hat{\bs{g}}$ has been recovered, even from a partial trait vector, all of the Fourier-based tools described earlier can be applied to $\hat{\bs{g}}$ to glean details about the trait, such as its modularity, local influence, total roughness, or other features.

Finally, we can predict the trait values for unobserved genomes by taking the inverse Fourier transform the recovered gene network, i.e., by substituting~$\hat{\bs{g}}$ in~(\ref{Eqn:t=Hg}). The predicted full trait~$\hat{\bs{t}}$ is shown in the bottom-left plot\,---\,compare this with the ground truth in Figure~\ref{Fig:Entacmaea_t_g_gLevel}. The goodness-of-fit is shown in the bottom-right, with $R^2 = 0.92$.
The predicted trait values match the ground truth fairly well, considering only $\K=200$, i.e., $2.44\%$ of the network's $N=8192$ interactions were used.

Although, the results from using compressive sensing are quite good in this example, they were accomplished with a subsample ratio $M/N = 10\%$, which is not an extremely low rate. This is probably connected to the fact that the trait is not of exponential size, in addition to possible noise in the measured data.

\subsubsection{A weighted Fourier transform}
It is worth noting that the literature mentions more than one transform to analyze epistasis~\cite{ContextDependenceMutations_Poelwijk2016, FourierTaylorFitnessLandscapes_Weinberger1991}. An alternative proposal to~(\ref{Eqn:g=Ht}) is to also weight the epistatic terms as a function of their level~\cite{ContextDependenceMutations_Poelwijk2016}:
\begin{equation} \label{Eqn:Alt_transform_VHt}
\bs{g}_\text{alt} \:=\: \bs{V}_{\!n}\bs{\Syl}_n\bs{t}
\end{equation}
where $\bs{V}_{\!n}$ is a diagonal weighting matrix defined recursively for $n\ge1$ by
$$
\bs{V}_{\!n} \,=\;
\left[
     \begin{array}{cc}
       \tfrac{1}{2}\bs{V}_{\!n-1} & \bs{0} \\[5pt]
       \bs{0} & -\bs{V}_{\!n-1} \\
     \end{array}
   \right]
$$
with $\bs{V}_0 = 1$. This transform appears to have its origins in a fixed, single-reference epistasis analysis influenced by traditional linear regression. There are certainly benefits to this formulation, however, as $\bs{V}_{\!n}$ endows the higher-level terms with exponentially more weight (i.e., powers of~$2$), the end result is that it tends to destroy the desired sparsity necessary for compressive sensing. For example, applying the transform~(\ref{Eqn:Alt_transform_VHt}) to the \emph{Entacmaea quadricolor} fluorescent protein trait~$\bs{t}$, we obtain the gene network~$\bs{g}_\text{alt}$ shown in the top of Figure~\ref{Fig:Alt_transform_VHt}. The bottom plot is just the level-ordered version of the same transform. This is a markedly dif\mbox{}ferent outcome from the non-weighted Fourier transform~(\ref{Eqn:g=Ht}) we advocate throughout this paper\,---\,contrast the extremely dense gene network representation of Figure~\ref{Fig:Alt_transform_VHt} with the extremely sparse version in Figure~\ref{Fig:Entacmaea_t_g_gLevel}. The bottom plot of Figure~\ref{Fig:Alt_transform_VHt} also shows how the higher levels are clearly weighted more strongly. This weighted epistasis transform fails to compress the trait into low levels, because it is not isometric (up to scale) to the trait data. In contrast, the plain Fourier transform~(\ref{Eqn:g=Ht}) facilitates compressive sensing due to the fortuitous features of the Sylvester-Hadamard matrix explained at the end of Section~\ref{Sec:Traits&theirNetworks}.
\begin{figure}[!tbh]
\centering
\includegraphics[scale=0.375, trim={0mm 10mm 0mm 8mm},clip] % {left, bottom, right, top}
{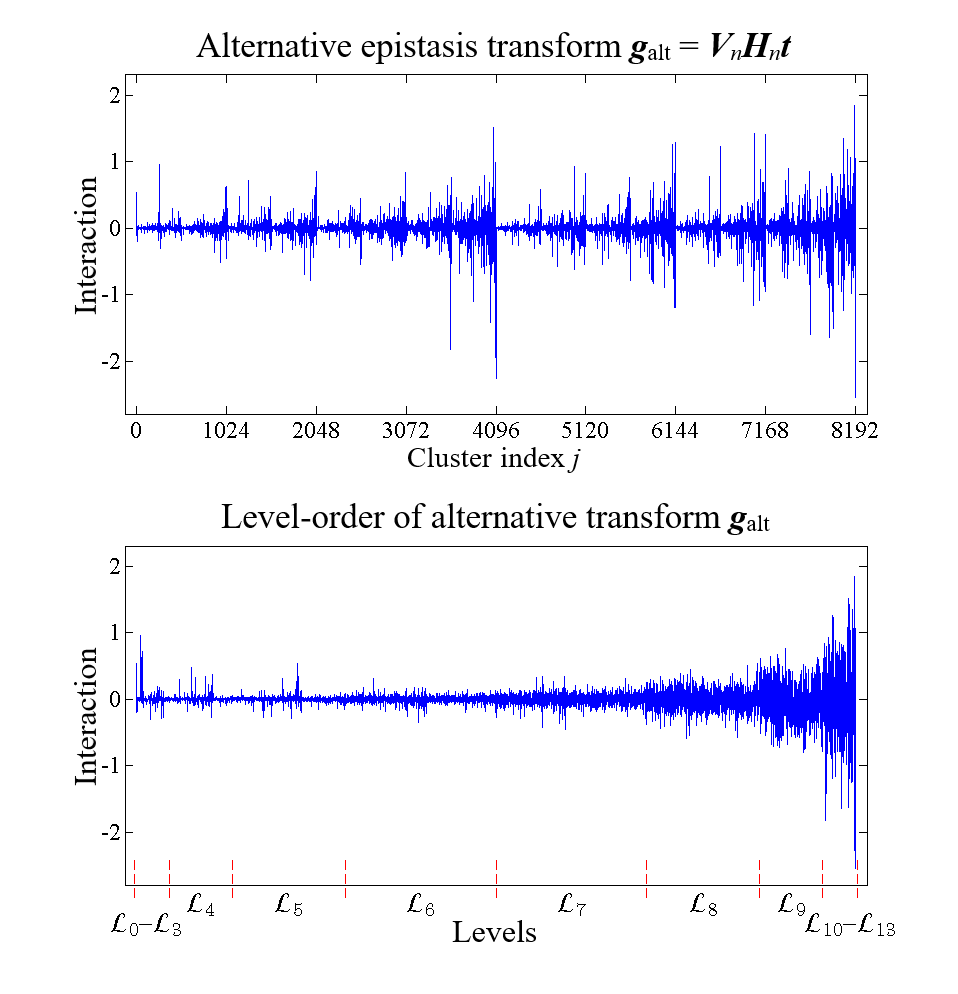}
\caption{(\textbf{Top}) Alternative transform~$\bs{g}_\text{alt}$ of the proteomic trait using~(\ref{Eqn:Alt_transform_VHt}). (\textbf{Bottom}) The level-ordered version of~$\bs{g}_\text{alt}$.
Compare these with the non-weighted Fourier transform~(\ref{Eqn:g=Ht}) in the plots on the right side of Fig.~\ref{Fig:Entacmaea_t_g_gLevel}.} \label{Fig:Alt_transform_VHt}
\end{figure}
\end{example}

% ==============================================================
% ==================================================================================
\section{Conclusion}
% ==================================================================================
% ==============================================================
In this paper, we define a spectrum of gene interactions that permit the analysis of quantitative traits as a sum of contributions from all possible sets of gene loci. The exponential expansion of combinations leads to the scale problem. We explore heuristic arguments that many traits may permit the application of compressive sensing to describe traits in terms of a low-level concentrated set of gene network interactions.

The uncertainty principle for the Fourier transform ensures that, whenever the gene network of a trait is suf\mbox{}ficiently concentrated, the ef\mbox{}fects in trait space are pervasive, so that a limited number of randomly chosen organisms will provide useful data. In ef\mbox{}fect, the ``needles in the haystack'' of the gene network cannot hide because they will poke all of the genomes. Much work still remains to test this theory empirically. This may involve recasting biological traits by a monotone transformation. But Poelwijk's proteomics example~\cite{LearningPatternEpistasis_Poelwijk2019}
shows the viability of this approach. The time may be ripe for further application of this technique. For instance, various amino acid substitutions either in the SARS-CoV-2 spike protein or in the ACE2 receptor may af\mbox{}fect their mutual binding strength~\cite{SARS-CoV-2SpikeMutation_Ozono2021, MutationsSARS-CoV-2SpikeProtein_Ortega2021, InteractionSpikeProtein_Othman2020} (taking binding strength as the relevant trait). Conveniently, the analysis need not be confined to just one protein: the amino acid substitutions can occur in both the spike protein and receptor. This application of Fourier techniques to proteomics might permit a rapid prediction of the features of a novel spike mutation spreading through genetically diverse populations.

The mathematical machinery of real-valued Boolean functions provides a language to describe fitness landscapes, and their properties such as roughness. This language permits one to formulate hypotheses about landscape features such as, (i) the density distribution of local roughness, (ii) the distribution of strong interactions by level. With empirical support confined to only one example, only heuristic arguments can be tentatively advanced, at present. Hence, we have concentrated in this paper on the possibility of actually detecting the gene network from observations of a small subset of genomes. Databanks of trait data for substantial and varied genomes are being collected and stored, as in the UK Biobank. The speed of computers enables more elaborate computation. Compressive sensing is a transparent alternative to ``black box'' machine learning technology.

Because we focused on establishing the computability of the network coef\mbox{}ficients, the gene network is defined in a way independent of the population distribution. But the population distribution can readily be incorporated into Boolean function analysis. The structure of a population is just another Boolean function, a distribution on the Boolean lattice of genomes. Hence, it has its own transform in network space. Averages for special populations are obtained by pointwise multiplication of this probability distribution and the trait. On the network side, this corresponds to a convolution multiplication of the transforms of trait and distribution. One could also analyze pleiotropy by looking at two or more separate traits in conjunction, creating ``vector-valued'' traits. This work is in progress and goes beyond the aims of this paper.

Knowledge of the gene network is valuable for two reasons. First, it provides insight as to the important interactions of the gene loci and overall gene function. Finding such interactions might help in designing drugs and other gene therapies, screening individuals for drug trials and warning of adverse interactions, analyzing disease susceptibility, and anticipating the ef\mbox{}fects of gene editing (e.g., CRISPR). Second, the gene network is a concise formulation predicting average trait values for unobserved organisms via the inverse Fourier transform (\ref{Eqn:t=Hg}). A consistent Fourier approach brings considerable unification and insight to the notions of trait terrain.

We successfully demonstrated a compressive sensing example with a subsample ratio of $M/N=10\%$ due to an approximate sparsity ratio of $\K/N = 2.44\%$; although this is small, it is not remarkable. We attribute this to the fact that the trait was governed by only $n=13$ loci. Based on the analysis in Section~\ref{Sec:Sparsity_fnc_n} we anticipate much more dramatic sparsity ratios $\K/N$ for larger $n$, and therefore much more impressive subsample ratios~$M/N$ on account of the linear relationship in~(\ref{Eqn:M_trait}). In fact, in the case of the sparsity~$\K$ scaling polynomially with~$n$, we see that~(\ref{Eqn:M_trait}) only adds an additional factor of~$n$ to the degree of the polynomial for the number of measurements~$M$, while the number of possible genotypes and interactions explodes to $N=2^n$.

In general, better performance and lower error may also be achieved by using custom recovery algorithms, as well as machine learning methods designed specifically for this application that exploit prior information. At the same time, there are significant challenges associated with memory, data communication, and processing of exponentially-large vectors in the reconstruction step. For very large~$n$, it may even be computationally intractable to implement~(\ref{Eqn:Basis_Pursuit}). In that case, a solution may be to only recover certain levels with the understanding that this will incur some residual error, as well as obviate the ability to implement fast Fourier transform methods.

%%\appendix
\section{Dedication}
Dedicated to the memory of physician-scientist John T.~Flynn, $1931$--$2019$.\\

\section{Acknowledgement}
M.A.H.~extends appreciation for generous hospitality from the \emph{Unemployed Philosopher's Guild} in  New York City, where some of this manuscript was written.

\bibliographystyle{elsarticle-num}


%\begin{thebibliography}{00}
%%% Text of bibliographic item
%
%\bibitem{}
%
%\end{thebibliography}

\end{document}